\documentclass[11pt,a4paper]{iopart}

\pdfoutput=1

\usepackage[utf8]{inputenc}
\usepackage{cite}
\usepackage{xcolor}
\usepackage{comment}

\usepackage{iopams}
\usepackage{braket}
\usepackage{bm}

\usepackage{graphicx}
\usepackage{subcaption}
\usepackage{tikz}
\usepackage{dcolumn}
\usepackage{multirow}

\makeatletter
\expandafter\let\csname equation*\endcsname\relax
\expandafter\let\csname endequation*\endcsname\relax
\makeatother

\usepackage{amsmath}

\usepackage{enumitem}

\usepackage{hyperref}
\usepackage{cleveref}

\hypersetup{
    colorlinks=true,
    linkcolor=blue,
    urlcolor=blue,
    citecolor=blue
}

\makeatletter

\renewcommand{\rmi}{\mathrm{i}}
\renewcommand{\rmd}{\mathrm{d}}
\renewcommand{\rme}{\mathrm{e}}

\newcommand{\appendixtocformat}{%
    \renewcommand{\l@section}
        {\@dottedtocline{1}{0em}{7em}}%
    \renewcommand{\l@subsection}
        {\@dottedtocline{2}{2em}{8.5em}}%
    \renewcommand{\l@subsubsection}
        {\@dottedtocline{3}{4em}{10.5em}}%
}

\makeatother

\newcommand{\proj}[1]{\ket{#1}\!\bra{#1}}
\newcommand{\comm}[2]{\left[#1,#2\right]}
\newcommand{\acomm}[2]{\left\{#1,#2\right\}}

\begin{document}
\title{Quantum resetting with memory}

\author{Gabriele de Mauro\footnote[7]{Corresponding author}}
\address{LPTMS, CNRS, Univ. Paris-Sud, Universit\'e Paris-Saclay, 91405 Orsay, France\\
gabriele.de-mauro@universite-paris-saclay.fr}

\author{Manas Kulkarni}
\address{International Centre for Theoretical Sciences, Tata Institute of Fundamental Research,
Bengaluru 560089, India\\
manas.kulkarni@icts.res.in }

\author{Satya N. Majumdar}
\address{LPTMS, CNRS, Univ.  Paris-Sud,  Universit\'e Paris-Saclay,  91405 Orsay,  France\\
satyanarayan.majumdar@cnrs.fr}

\date{\today}

\begin{abstract}
We introduce a quantum stochastic resetting protocol with uniform memory, in which each resetting event returns the system to a state visited at a time chosen uniformly from its entire history. The resulting dynamics is nonunitary, non-Markovian and a direct quantum generalization of the classical preferential relocation model. Working in the energy eigenbasis, we derive the exact evolution of every density-matrix element for an arbitrary time-independent Hamiltonian and show that the Hamiltonian enters the dynamics only through the corresponding Bohr frequencies. This leads to a natural distinction between two classes of quantum systems: gapped and gapless. In \emph{gapped systems} (systems with a discrete energy spectrum), while the diagonal elements remain unchanged, the off-diagonal elements of the density matrix in the energy eigenbasis decay algebraically with a continuously varying exponent and with an amplitude that oscillates periodically in $\log t$. The system therefore approaches a stationary state that is independent of the resetting rate and retains a strong memory of the initial state. In \emph{gapless systems} (systems with a continuous energy spectrum), arbitrarily small Bohr frequencies prevent stationarity. 
Instead, the position distribution spreads on the universal (ultra-slow) scale $\log(rt)/r$, independently of the initial state and of the details of the Hamiltonian.
We illustrate these results with a two-level system, a harmonic oscillator, and a free quantum particle, and contrast them with their classical counterparts.
\end{abstract}

\maketitle


\section{Introduction}

Stochastic resetting~\cite{EM2011,EM2011JPA,EMSch2020,GuptaJ2022,KR2024,KMSGRRH2026} consists of interrupting a dynamical process at random times and returning the system to a prescribed state. By introducing an additional time scale, it can qualitatively reshape the underlying dynamics, with important consequences for first-passage and search properties. These effects have been studied in diffusive target-search problems~\cite{EM2011,EM2011JPA,R2016,NagarGupta2016,PR2017,CS2018,DeBSNMGS2022}, under time-dependent resetting~\cite{PKE2016}, in spatially dependent or heterogeneous environments~\cite{P2018,GPPM2023,VBGN2026}, and in the presence of partially absorbing targets~\cite{WEM2013,SB2021}. 
Another important protocol is threshold resetting \cite{DRR2020,BMSP2025_ts}, in which a reset is triggered when the process reaches a prescribed threshold rather than by an external Poisson clock. Related developments include collective searches~\cite{BdBR2026,BMS2023}, lattice systems~\cite{Christophorov,BGlattice,HM2025}, and adaptive navigation strategies~\cite{DeBMoriOptcontr,MoriMahad2025,DKMS2026}. 
Besides providing an efficient search strategy, stochastic resetting also has another important aspect, namely that it drives a system to a nonequilibrium stationary state (NESS)~\cite{EM2011,EM2011JPA,PKR2022,GMS2014,BKP2019,MMS2020,CZ2024,AcharyaMajumderGupta2025,AN2026}. Moreover, a common stochastic driving can induce strong correlations between otherwise independent degrees of freedom. Such dynamically emergent correlations (DEC) have been investigated in Brownian gases under different resetting protocols~\cite{Biroli2023,deMauro2026_NonPoisson,BMS2026_fpg,deMauro2026Conf,BM2026a,BMS2025,VilkAssafMeerson2022,MeersonVilk2026,Vilk2026}, in systems driven by fluctuating common environments~\cite{BKMS2024,SabM2024,MMS2025}, and within more general theoretical frameworks~\cite{BLMS2024,Galla2026,Olsen2026}. The same mechanism has also been extended to batch resetting events, where only a fraction of the particles is reset while the others continue to evolve via their reset-free dynamics~\cite{deMauro2026}. 
Experimental realizations have been achieved in colloidal systems using optical trapping techniques, both for single particles~\cite{Besga2020,Faisant2021,Tal-Friedman2020,GinotBechinger2025} and for many-particle systems~\cite{Vatash2025,Biroli2026exp}. For a recent perspective article on DEC see Ref. \cite{MS2026_DEC}.

More recently, stochastic resetting has been extended to quantum systems, with studies focusing on the resulting dynamics and stationary states~\cite{MSM2018,RTLG2018,PCML2021,WaldBottcher2021,DasDattaguptaGupta2022,AcharyaGupta2023,SVH2023,PMCMR2024,WaldYaoPlatiniHooleyCarollo2025}, open quantum systems \cite{PerfettoCarolloLesanovsky2022,CarolloLesanovskyGarrahan2024,SolankiLesanovskyPerfetto2025}, first-detection and quantum-search problems~\cite{KM2023Detection,YB2023,YinWangBarkai2024,YWTB2025,KRMKBG2026}, and many-body correlations and entanglement~\cite{MCPL2022,KM2023,TDFS2022,KMS2025,SLP2025,GKP2026,Ghosh2026,MurauerTornowPerfetto2026}.

In most of the models mentioned above, classical or quantum, the resetting protocol is memory-less and the system is returned to a fixed initial configuration or state. However, even in this case, memory effects can arise from coupling to an environment that is not reset and retains information about the past, as recently studied for a particle in a viscoelastic bath~\cite{Santra2026}. A different class of classical models, where memory enters directly into the resetting protocol, has also been studied, leading to interesting and nontrivial memory effects. These include, in particular, the ``preferential relocation" model of a random walk, in which the walker is relocated at a constant rate $r$ to
a position visited at a past time chosen uniformly from its entire history \cite{BSS2014}. This is also known as the ``monkey walk", since certain species of capuchin monkeys follow this resetting pattern during their foraging period \cite{BSS2014}. This model has also been studied in the mathematical literature, where several rigorous results were obtained \cite{MUB2019}.
These studies were extended further to more general memory kernels \cite{BRC2014} and to
generic random-walk settings \cite{BP2016}, including L\'evy processes.
The preferential relocation model in the presence of a spatial impurity was
studied in Refs.~\cite{FBGM2017,BFGM2019,Sepehrinia2025,Sepehrinia2026,ArmfieldMailler2026}, where the interplay between
memory and spatial heterogeneity was shown to produce an Anderson-like
localization transition. A mini-review of random walks with memory-induced
relocations can be found in Ref.~\cite{MCM2019}.
A continuous-space formulation for Brownian
diffusion was introduced in Ref.~\cite{BEM2017}, together with a general
memory kernel $K(\tau,t)$ that interpolates between Poissonian resetting to
the initial position and the preferential relocation protocol. Related
models have also been studied for active particles~\cite{BM2024a}, sluggish
random walkers~\cite{BM2026b}, confined diffusion~\cite{BM2024b,BEM2025},
and many-particle systems~\cite{BM2026a}. Depending on the temporal bias of
the memory kernel, the dynamics can exhibit conventional diffusion,
anomalous diffusion, or ultraslow spreading~\cite{BEM2017}. 

This naturally raises the question: what is the effect of memory on the resetting protocol for quantum systems?
Incorporating these memory effects in quantum systems is nontrivial. Unlike a classical trajectory, a quantum system is described by a density matrix that contains populations (diagonal elements of the density matrix) and coherences (off-diagonal elements), and a reset to a previously visited state must retain the corresponding quantum information. The resulting evolution combines coherent dynamics, stochastic resets, and temporal nonlocality, and is therefore generally nonunitary and non-Markovian~\cite{CarolloWald2026}. Moreover, because the reset state depends explicitly on the preceding evolution, the conventional renewal formulation used for fixed-state quantum resetting~\cite{MSM2018} is no longer directly applicable.

In this work, we introduce a quantum resetting protocol with memory in which, at every resetting event, the system is returned to a state visited at an earlier time selected uniformly from its entire history. Thus, all previous times are chosen with equal probability. This is the quantum analog of the classical preferential relocation model. We formulate the corresponding evolution equation and develop a general analytical framework to study the resulting nonunitary and non-Markovian dynamics, contrasting it with conventional quantum resetting to the initial state~\cite{EM2011,EM2011JPA,MSM2018}.
Working in the energy eigenbasis, we derive the exact evolution of each density-matrix element for an arbitrary time-independent Hamiltonian and show that the microscopic details of the Hamiltonian enter only through the corresponding Bohr frequencies. We then investigate the consequences of this very general result for two broad classes of quantum systems. For \emph{gapped quantum systems}, more specifically, systems with a discrete energy spectrum, the relevant nonzero Bohr frequencies are bounded away from zero, leading to the algebraic suppression of coherences and relaxation towards a stationary state. We then illustrate this behavior using two concrete examples: a quantum two-level system and a quantum harmonic oscillator. For \emph{gapless quantum systems}, specifically systems with a continuous energy spectrum, arbitrarily small Bohr frequencies are instead present, producing qualitatively different long-time dynamics. We illustrate this regime using a freely propagating quantum particle with a general dispersion relation, whose reset-free evolution exhibits unbounded dispersive spreading.

This classification exposes a qualitative distinction in the long-time behavior. For gapped systems, uniform memory resetting drives the system towards a stationary state through an ultraslow power-law relaxation, in sharp contrast with the exponential approach typically observed. This algebraic decay is further accompanied by very slow oscillations, whose phase grows only logarithmically in time. Remarkably, the stationary state reached at long times coincides with the time average of the reset-free unitary dynamics and is therefore completely independent of the resetting rate $r$, while retaining a very strong memory of the initial condition. For gapless systems, by contrast, no stationary state is reached. Instead, the spatial extent continues to grow logarithmically in time. This $\log t$ growth is universal and determined solely by the presence of memory, whereas the asymptotic shape of the spreading distribution depends on both the initial state and the specific dispersion relation, and therefore also retains a very strong memory of the initial condition. Thus, in both cases, the dynamics preserves a strong memory of the initial state, while the existence of a spectral gap determines whether the memory-induced ultraslow dynamics culminates in stationarity or in persistent spreading.

Despite this sharp distinction, both categories exhibit a common suppression of the intrinsic quantum dynamics. Under stochastic resetting with uniform memory, the continually expanding history progressively reduces the relative weight of newly generated states, leading to ultraslow evolution governed primarily by the temporal structure of the memory kernel rather than by the microscopic details of the Hamiltonian. This behavior contrasts sharply with conventional resetting, which typically produces an exponentially fast approach to a NESS. Quantum resetting with memory therefore provides a general framework for generating and controlling slow, non-Markovian dynamical regimes in both gapped and gapless quantum systems.

The paper is organized as follows. In Sec.~\ref{sec:protocol}, we formulate the general protocol for quantum resetting with memory. We then discuss gapped quantum systems in Sec.~\ref{sec:gap}, using the two-level system and the quantum harmonic oscillator as representative examples. In Sec.~\ref{sec:gapless}, we discuss gapless quantum systems, illustrated by the free quantum particle. Conclusions and an outlook are presented in Sec.~\ref{sec:conc}, while technical details are relegated to the appendices.

\section{Quantum resetting protocol with memory}
\label{sec:protocol}

In this section, we formulate a general framework for quantum stochastic
resetting with memory. We first review the reset-free unitary dynamics and
conventional Poissonian resetting to a fixed state, and then extend the
corresponding evolution equation to a protocol in which the reset state is
sampled from the system's own history.

We consider a closed quantum system governed by a time-independent
Hamiltonian $\hat H$. At $t=0$, the system is prepared in a general density
matrix $\hat\rho(0)$, which may describe either a pure or a mixed state.
Throughout this work, we set $\hbar=1$. In the absence of resetting, the
density matrix $\hat{\rho}_0(t)$ evolves unitarily as \cite{FHS2010}
\begin{equation}
\label{eq:density_matrix_timet}
\hat{\rho}_0(t)
=
\rme^{-\rmi\hat Ht}
\hat{\rho}(0)
\rme^{\rmi\hat Ht}.
\end{equation}
Throughout the paper, the subscript $0$ denotes the absence of resetting.
Equivalently, Eq.~\eqref{eq:density_matrix_timet} satisfies the von Neumann equation
\begin{equation}
\frac{\rmd\hat{\rho}_0(t)}{\rmd t}
=
-\rmi\left[\hat H,\hat{\rho}_0(t)\right],
\end{equation}
with initial condition $\hat{\rho}(0)$.

We now introduce stochastic resetting events occurring according to a Poisson
process with constant rate $r$. The probability that no reset occurs during
a time interval of duration $t$ is $\rme^{-rt}$, while the inter reset interval
$\tau$ between consecutive resets is distributed as
$R(\tau)=r\,\rme^{-r\tau}$. Starting from $\hat\rho(0)$, the dynamics is
defined by the following protocol:
\begin{enumerate}[label=(\roman*),ref=(\roman*)]
    \item\label{step:unitary_evolution}
    An inter reset interval (say $\tau_1$) is drawn from the exponential distribution
    $R(\tau)=r\,\rme^{-r\tau}$. During this interval, the system evolves
    unitarily according to Eq.~\eqref{eq:density_matrix_timet}.

    \item\label{step:resetting}
    At time $\tau_1$, a resetting event occurs and the system is
    instantaneously returned to its initial state $\hat{\rho}(0)$.
\end{enumerate}
After the reset, a new inter reset interval (say $\tau_2$) is independently drawn from the
same distribution $R(\tau)$. The system then evolves unitarily for a duration $\tau_2$
before being reset again to $\hat{\rho}(0)$. This sequence of unitary
evolution intervals and instantaneous resetting events is repeated
indefinitely.

We denote by $\hat{\rho}_r(t)$ the density matrix averaged over all possible
realizations of the resetting process, where the subscript $r$ indicates a resetting rate $r$. 
The expectation value of a generic observable $\hat O$ is then given by $\left\langle \hat O(t)\right\rangle_r = \operatorname{Tr}\left[ \hat{\rho}_r(t)\hat O \right]$.
Since the previous history of the process becomes irrelevant once a reset event occurs, the density matrix $\hat{\rho}_r(t)$ satisfies the renewal equation \cite{EMSch2020,MSM2018}
\begin{equation}\label{eq:renewal_standard_resetting}
\hat{\rho}_r(t)=\rme^{-rt}\hat{\rho}_0(t)+
r\int_0^t \rmd\tau\,
\rme^{-r\tau}\hat{\rho}_0(\tau).
\end{equation}
The first term accounts for realizations in which no resetting event occurs up to time $t$. This happens with probability $\rme^{-rt}$, and the system therefore evolves unitarily from its initial state for the entire time interval $t$, resulting in the reset-free density matrix $\hat{\rho}_0(t)$. The second term accounts for realizations in which at least one resetting event has occurred. In this term, $\tau$ denotes the time elapsed since the most recent reset. Because that reset returns the system exactly to its initial state, the evolution before the most recent reset is irrelevant. The system subsequently evolves freely for a time $\tau$, reaching the state $\hat{\rho}_0(\tau)$. The factor $r\,\rme^{-r\tau}\rmd\tau$ gives the probability that the most recent reset occurred between $t-\tau$ and $t-\tau+\rmd\tau$, with no further resets during the following interval of duration $\tau$. Integrating over $\tau\in[0,t]$ therefore averages over all possible times elapsed since the most recent resetting event. 

Equation~\eqref{eq:renewal_standard_resetting} is equivalent to the following ``Master equation" [see \ref{app:equiv}]:
\begin{equation}\label{eq:master_standard_resetting}
\frac{\rmd\hat{\rho}_r(t)}{\rmd t}
=
-\rmi\left[\hat H,\hat{\rho}_r(t)\right]
-r\hat{\rho}_r(t)
+r\hat{\rho}(0).
\end{equation}
The first term describes the unitary evolution generated by $\hat H$, the second term accounts for the loss of probability from the current state due to resetting, and the last term reinjects this probability into the fixed reset state $\hat{\rho}(0)$.
Equation~\eqref{eq:renewal_standard_resetting} provides the explicit solution to Eq.~\eqref{eq:master_standard_resetting} and is generally the most convenient formulation whenever a renewal description is available. For this reason, previous studies of quantum stochastic resetting have typically relied directly on the renewal equation~\eqref{eq:renewal_standard_resetting}. In the present work, however, it is useful to introduce the equivalent differential formulation in Eq.~\eqref{eq:master_standard_resetting}. As we will see below, this formulation can be generalized more naturally to situations in which a simple renewal decomposition is no longer available, as occurs when memory effects are introduced. In \ref{subsec:lindblad} we also show how Eq.~\eqref{eq:master_standard_resetting} can be expressed exactly in Lindblad form.

We now use the differential formulation given in Eq. \eqref{eq:master_standard_resetting} to generalize the resetting protocol to the case in which the reset state depends on the system’s previous evolution. The protocol again begins with step \ref{step:unitary_evolution}, while step \ref{step:resetting} is slightly modified: the system is no longer returned to a fixed state. Instead, when a resetting event occurs at time $t$, a past time $\tau$ is selected uniformly in $[0,t]$. Averaging over all realizations of this stochastic resetting protocol gives the following evolution equation (see~\ref{app:uniform_memory_derivation} for a derivation):
\begin{equation}\label{eq:uniform_memory}
\frac{\rmd\hat{\rho}_r(t)}{\rmd t}
=
-\rmi\left[\hat H,\hat{\rho}_r(t)\right]
-r\hat{\rho}_r(t)
+
\frac{r}{t}
\int_0^t \rmd\tau\,\hat{\rho}_r(\tau).
\end{equation}
The first two terms in Eq.~\eqref{eq:uniform_memory} are identical to those in Eq.~\eqref{eq:master_standard_resetting} and retain the same physical
interpretation. The first term describes the unitary evolution generated by
the Hamiltonian $\hat H$, while the second accounts for the loss of
probability from the state occupied at time $t$ due to resetting events.
The last term generalizes the reinjection term
$r\hat{\rho}(0)$ appearing in
Eq.~\eqref{eq:master_standard_resetting}. In conventional resetting, the system is always returned to the fixed initial state $\hat{\rho}(0)$. In the
uniform memory protocol, instead, a past time $\tau$ is sampled uniformly in $[0,t]$, and the system is reset to
the state previously occupied at that time, described by $\hat{\rho}_r(\tau)$. The
quantity $\rmd\tau/t$ in the last term of Eq. \eqref{eq:uniform_memory} is the probability of selecting a past time in the interval $[\tau,\tau+\rmd\tau]$. The integral therefore averages the reset state over all possible past times, while the prefactor $r$ accounts for the rate at which resetting events occur.
Since the reinjection term depends on the density matrix at all previous times $\tau\leq t$, the resulting evolution is nonlocal in time. Consequently, the dynamics is generally non-Markovian and does not admit the simple renewal representation of Eq.~\eqref{eq:renewal_standard_resetting}. The main goal of this work is to study Eq.~\eqref{eq:uniform_memory} for two very general classes of quantum systems: \emph{gapped systems}, namely systems whose Hamiltonian has a discrete spectrum, and \emph{gapless systems}, namely systems whose Hamiltonian has a continuous spectrum.

Equation~\eqref{eq:uniform_memory} is the quantum counterpart of the
Fokker--Planck equation for the classical position distribution $P_r(x,t)$
under uniform memory resetting~\cite{BSS2014,BRC2014,BP2016,BEM2017,
FBGM2017,BFGM2019,MCM2019,MUB2019,BM2024a,BM2024b,BEM2025,BM2026a,
BM2026b}, with the density matrix $\hat\rho_r(t)$ playing the role of the
classical probability distribution and the reset-free classical dynamics
replaced by the unitary evolution generated by $\hat H$. Classical resetting
with memory has been extensively studied because the repeated sampling of
past configurations can strongly modify the long-time dynamics, producing
anomalous diffusion, ultraslow spreading, and unusually slow relaxation
towards stationary states. In particular, a Brownian particle diffusing on
the infinite line under uniform memory resetting was studied in
Ref.~\cite{BEM2017}. In this case, no stationary state is reached and the
position distribution remains asymptotically Gaussian, but with a characteristic width growing as $\sqrt{\log t}$. This system can be regarded as the
classical analogue of the gapless quantum systems considered here, since the
corresponding Fokker--Planck operator has a continuous spectrum extending
down to zero. By contrast, classical Brownian particles evolving in a
confining potential $V(x)$ were studied in
Refs.~\cite{BM2024b} with the same uniform memory kernel. Remarkably, despite the history-dependent
resetting dynamics, the probability distribution approaches the equilibrium
Gibbs--Boltzmann state at long times, independently of the initial condition.
The relaxation towards this stationary state is, however, extremely slow
and occurs algebraically rather than exponentially, with an exponent that depends on the resetting rate $r$ and the smallest nonzero eigenvalue of the Fokker--Planck operator. These confined systems
provide the natural classical counterparts of the gapped quantum systems
studied here, since their Fokker--Planck operator has a discrete spectrum
with a finite gap above its stationary mode. More generally, Ref.~\cite{BEM2017} studied the unconfined Brownian particle for a generic memory kernel $K(\tau,t)$, with uniform memory corresponding to the particular choice $K(\tau,t)=1/t$. Similarly, Ref.~\cite{BEM2025} extended the analysis of confined diffusion to a broad class of memory kernels, showing how the sampling of the past affects both the stationary state and the relaxation towards it. In the present work, however, we focus exclusively on the uniform memory kernel and leave the extension to more general memory kernels as an interesting open problem.

As we will see, the classical and quantum problems share the same basic
spectral distinction. In both cases, a discrete spectrum with a finite gap
leads to an ultraslow power-law relaxation towards a stationary state,
whereas a continuous spectrum extending down to zero prevents stationarity
and produces persistent ultraslow spreading. The long-time behavior,
however, differs in crucial ways. For confined classical systems, the
stationary state is always the equilibrium Gibbs--Boltzmann distribution~\cite{BM2024b} and it
is therefore independent of the initial condition, while in the gapped quantum case it
retains a strong memory of the initial state. In particular, for a nondegenerate spectrum, the stationary density matrix is given by the initial populations in the energy eigenbasis,
\begin{equation}\label{eq:diag_statsate_intro}
\hat\rho_r^{\rm st}
=
\sum_n
\rho_{nn}(0)\ket{n}\bra{n}.
\end{equation}
Here and throughout the paper, we denote the matrix elements of the initial density matrix by $\rho_{mn}(0)$, without specifying the subscript $0$ or $r$ (indicating the absence or presence of resetting, respectively), since the initial state is fixed and chosen to be the same for both dynamics.
Similarly, for an unconfined classical particle, the distribution remains
asymptotically Gaussian and its width grows as $\sqrt{\log t}$ \cite{BEM2017}, whereas in
the gapless quantum case the width grows as $\log t$. Moreover, the
asymptotic quantum distribution again retains a strong memory of the
microscopic dynamics, since its shape depends on both the initial state and
the specific dispersion relation. These differences originate from the
coherent unitary dynamics and therefore represent genuinely quantum effects.

Having established the classical--quantum correspondence and anticipated the
central role played by the spectral properties of the underlying dynamics, we
now turn to the exact solution of Eq.~\eqref{eq:uniform_memory}. The energy
eigenbasis provides the natural framework for this analysis, since it
diagonalizes the unitary part of the evolution and makes the dependence on
the energy spectrum explicit. Without yet assuming whether the spectrum is discrete or continuous, we denote its energy eigenstates generically by $\hat H\ket{m}=E_m\ket{m}$ and define $\rho_{r,mn}(t)=\bra{m}\hat{\rho}_r(t)\ket{n}$. Projecting
Eq.~\eqref{eq:uniform_memory} between $\bra{m}$ and $\ket{n}$, we obtain
\begin{equation}\label{eq:general_memory_energy_basis}
\frac{\rmd\rho_{r,mn}(t)}{\rmd t}
=
-\left[r+\rmi(E_m-E_n)\right]\rho_{r,mn}(t)
+
\frac{r}{t}
\int_0^t\rmd\tau\,\rho_{r,mn}(\tau).
\end{equation}
Equation~\eqref{eq:general_memory_energy_basis} shows that the different
matrix elements of $\hat{\rho}_r(t)$ evolve independently in the energy
basis. The Hamiltonian enters their evolution only through the corresponding
Bohr frequency $\omega_{mn}\equiv E_m-E_n$. More precisely, since Eq.~\eqref{eq:general_memory_energy_basis}
is linear and homogeneous in $\rho_{r,mn}(t)$,
the evolution of every matrix element can be written as
\begin{equation}\label{eq:general_memory_factorization}
\rho_{r,mn}(t)
=
\rho_{mn}(0)
f\left(\omega_{mn},t\right),
\end{equation}
where the same scalar function $f(\omega,t)$, evaluated at the corresponding Bohr frequency, governs the time dependence of every matrix element. It therefore remains to determine $f(\omega,t)$. 
Substituting Eq.~\eqref{eq:general_memory_factorization} into Eq.~\eqref{eq:general_memory_energy_basis}, we obtain
\begin{equation}
\label{eq:general_memory_scalar_function}
\frac{\partial f(\omega,t)}{\partial t}=-\left(r+\rmi\omega\right)f(\omega,t)+\frac{r}{t}
\int_0^t\rmd\tau\,f(\omega,\tau).
\end{equation}
Multiplying Eq.~\eqref{eq:general_memory_scalar_function} by $t$ and differentiating with respect to time eliminates the memory integral and yields
\begin{equation}
\label{eq:general_memory_scalar_second_order}
t \frac{\partial^2 f(\omega,t)}{\partial t^2}+\left[1+\left(r+\rmi\omega\right)t\right]
\frac{\partial f(\omega,t)}{\partial t}+\rmi\omega f(\omega,t)=0.
\end{equation}
Because this is a second-order equation, it requires two initial conditions. The first follows directly from Eq.~\eqref{eq:general_memory_factorization}, namely $f(\omega,0)=1$. The second is obtained by taking the limit $t\to0$ in Eq.~\eqref{eq:general_memory_scalar_function}. Since $\lim_{t\to0}t^{-1}\int_0^t\rmd\tau\,f(\omega,\tau)=f(\omega,0)$, one finds $\left.\frac{\partial f(\omega,t)}{\partial t}\right|_{t=0}=-\rmi\omega$.
We now introduce $\chi(\omega)=\rmi\omega/(r+\rmi\omega)$. Separating its real and
imaginary parts as
$\chi(\omega)=\chi_{\rm R}(\omega)+\rmi\chi_{\rm I}(\omega)$, we have $\chi_{\rm R}(\omega)=\frac{\omega^2}{r^2+\omega^2}$ and $\chi_{\rm I}(\omega)=\frac{r\omega}{r^2+\omega^2}$.
Defining $u=-(r+\rmi\omega)t$ and using
$\frac{\partial}{\partial t}=-(r+\rmi\omega)\frac{\partial}{\partial u}$,
Eq.~\eqref{eq:general_memory_scalar_second_order} becomes
\begin{equation}\label{eq:general_memory_kummer_equation}
u\frac{\partial^2 f}{\partial u^2}+(1-u)\frac{\partial f}{\partial u}-\chi(\omega)f=0.
\end{equation}
Eq. \eqref{eq:general_memory_kummer_equation} is Kummer's confluent hypergeometric equation (see Eq. 13.2.1 of
Ref.~\cite{NIST}).
The general solution is a linear combination of $M(\chi;1;u)$ and $U(\chi;1;u)$, respectively known as Kummer's confluent hypergeometric function of the first kind and Tricomi's confluent hypergeometric function of the second kind.
Since $U(\chi,1,u)$ is singular at $u=0$ for generic
$\chi$, regularity at the initial time excludes this solution. Using
$M(\chi,1,0)=1$, the condition $f(\omega,0)=1$ then gives
\begin{equation}\label{eq:general_memory_kummer_solution}
f(\omega,t)
=
M\left(
\frac{\rmi\omega}{r+\rmi\omega};
1;
-\left(r+\rmi\omega\right)t
\right).
\end{equation}
The Kummer function $M(a;b;u)$ is defined by the series expansion~\cite{NIST}
\begin{equation}
M(a;b;u)
=
\sum_{k=0}^{\infty}
\frac{(a)_k}{(b)_k}
\frac{u^k}{k!},
\end{equation}
where $(a)_k=a(a+1)\cdots(a+k-1)$ is the rising factorial, with $(a)_0=1$.
From this expansion, we have $M(a;b;0)=1$ and
$\left. \frac{\partial M(a;b;u)}{\partial u}\right|_{u=0}=a/b$.
The latter property ensures that the solution also satisfies the second initial condition,
$\left.\frac{\partial f(\omega,t)}{\partial t}\right|_{t=0}
=-(r+\rmi\omega)\chi(\omega)=-\rmi\omega$.
For $\omega=0$, Eq.~\eqref{eq:general_memory_scalar_function} directly
gives the constant solution $f(0,t)=1$.
Consequently, the complete density matrix in the energy basis is
\begin{equation}\label{eq:general_memory_density_matrix_solution}
\rho_{r,mn}(t)
=
\rho_{mn}(0)
M\left(
\frac{\rmi\omega_{mn}}{r+\rmi\omega_{mn}};
1;
-\left(r+\rmi\omega_{mn}\right)t
\right).
\end{equation}
For the diagonal elements, $\omega_{mm}=0$, and therefore $\rho_{r,mm}(t)=\rho_{mm}(0)$.
Hence, the populations (diagonal elements) in the energy eigenbasis remain equal to their
initial values at all times. More generally, whenever two energy eigenstates
are degenerate, $E_m=E_n$, one has $\omega_{mn}=0$ and therefore $\rho_{r,mn}(t)=\rho_{mn}(0)$.
Thus, uniform memory resetting preserves both the energy populations and
the coherences within each degenerate energy eigenspace. The nontrivial
dynamics is entirely carried by matrix elements connecting states with
different energies.
Equation~\eqref{eq:general_memory_density_matrix_solution} is the central result of this work. It is completely
general and it applies to any time-independent Hamiltonian, irrespective of
whether its spectrum is gapped or gapless. Remarkably, the details of the reset-free unitary dynamics enter only through the Bohr frequencies $\omega_{mn}$.

We now analyze the general long-time behavior of Eq. \eqref{eq:general_memory_density_matrix_solution}. For a fixed nonzero $\omega$, the large-argument asymptotic expansion of Kummer's function gives
[see Eq.~(13.7.2) of Ref.~\cite{NIST}]
\begin{equation}\label{eq:protocol_f_asymptotic}
f(\omega,t)
\approx
\frac{
\left[
\left(r+\rmi\omega\right)t
\right]^{-\chi(\omega)}
}{
\Gamma\left(1-\chi(\omega)\right)
},
\qquad
t\to\infty,
\end{equation}
where we recall the definition of $\chi(\omega)=\rmi\omega/(r+\rmi\omega)=\chi_{\rm R}(\omega)+\rmi\chi_{\rm I}(\omega)$ given above Eq.~\eqref{eq:general_memory_kummer_equation}. 
We also write
$r+\rmi\omega=\sqrt{r^2+\omega^2}\,\rme^{\rmi\beta(\omega)}$, where
we define the phase $\beta(\omega)=\arctan(\omega/r)$. Equation~\eqref{eq:protocol_f_asymptotic}
can then be expressed as
\begin{equation}\label{eq:protocol_f_asymptotic_amplitude_phase}
f(\omega,t)
\approx
C(\omega)
\left[
\sqrt{r^2+\omega^2}\,t
\right]^{-\chi_{\rm R}(\omega)}
\rme^{-\rmi\left[
\chi_{\rm I}(\omega)
\log\left(
\sqrt{r^2+\omega^2}\,t
\right)
+
\varphi(\omega)
\right]},
\end{equation}
where
\begin{equation}\label{eq:protocol_f_amplitude}
C(\omega)
=
\frac{
\rme^{\chi_{\rm I}(\omega)\beta(\omega)}
}{
\left|
\Gamma\left(
1-\chi_{\rm R}(\omega)-\rmi\chi_{\rm I}(\omega)
\right)
\right|
}
\end{equation}
and
\begin{equation}\label{eq:protocol_f_phase}
\varphi(\omega)
=
\chi_{\rm R}(\omega)\beta(\omega)
+
\arg\Gamma\left(
1-\chi_{\rm R}(\omega)-\rmi\chi_{\rm I}(\omega)
\right).
\end{equation}
Consequently, the modulus of $f(\omega,t)$ decays algebraically as
$\left|f(\omega,t)\right|= O\left(t^{-\frac{\omega^2}{r^2+\omega^2}}\right)$,
while its phase oscillates periodically as a function of $\log t$, with logarithmic angular frequency
$\chi_{\rm I}(\omega)=r\omega/(r^2+\omega^2)$.
In particular, $f(\omega,t)\to0$ as $t\to\infty$ for every fixed
$\omega\neq0$, since
$0<\chi_{\rm R}(\omega)=\omega^2/(r^2+\omega^2)<1$. Therefore, every
density-matrix element connecting states with different energies vanishes
algebraically at long times:
\begin{equation}\label{eq:protocol_density_matrix_asymptotic}
\rho_{r,mn}(t)
\approx
\rho_{mn}(0)\,
C(\omega_{mn})
\left[
\sqrt{r^2+\omega_{mn}^2}\,t
\right]^{-\chi_{\rm R}(\omega_{mn})}
\rme^{-\rmi\left[
\chi_{\rm I}(\omega_{mn})
\log\left(
\sqrt{r^2+\omega_{mn}^2}\,t
\right)
+
\varphi(\omega_{mn})
\right]},
\end{equation}
for $E_m\neq E_n$. The last term in Eq. \eqref{eq:protocol_density_matrix_asymptotic} has unit modulus and a phase that grows only logarithmically in time, giving rise to extremely slow oscillations that are periodic in $\log t$.
For the diagonal elements, instead, the Bohr frequency vanishes, $\omega_{mm}=0$. 
The exact solution $f(0,t)=1$ therefore gives
\begin{equation}
\label{eq:constant_diag}
\rho_{r,mm}(t)=\rho_{mm}(0),
\end{equation}
for all times $t$. Hence, the populations in the energy eigenbasis remain unchanged throughout the evolution. Together with
Eq.~\eqref{eq:protocol_density_matrix_asymptotic}, Eq. \eqref{eq:constant_diag} shows that, whenever
the relevant nonzero Bohr frequencies are bounded away from zero, all
coherences between states with different energies vanish algebraically at
long times, while the populations remain fixed. The system consequently approaches the stationary state given in Eq.~\eqref{eq:diag_statsate_intro}.

This dephasing (loss of coherence) mechanism is reminiscent of dephasing Lindbladian terms in an open quantum system~\cite{GKSL1976,Lindblad1976,BreuerPetruccione2006,Carmichael2002}. However, the crucial difference is that the Lindbladian terms typically cause dephasing at an exponentially fast rate. In our case, the dephasing occurs because the system randomly revisits its own history, and it occurs algebraically slowly. Interestingly, while conventional quantum resetting can be expressed in Lindblad form (see~\ref{subsec:lindblad}), the presence of memory makes our dynamics non-Markovian, preventing, in general, such a description in terms of a time-independent Lindblad generator.

\section{Gapped Quantum Systems}
\label{sec:gap}

In this section, we discuss the implications of the general solution in
Eq.~\eqref{eq:general_memory_density_matrix_solution} for a gapped quantum
system. We consider a system described by a time-independent Hamiltonian with
a discrete and nondegenerate energy spectrum, $\hat H\ket{n}=E_n\ket{n},$ with $E_m\neq E_n$ for $m\neq n.$
We further assume that the nonzero Bohr frequencies
$\omega_{mn}\equiv E_m-E_n$ relevant to the dynamics are bounded away from zero. Namely, there exists a finite frequency gap
\begin{equation}
\label{eq:gapped_frequency_gap}
\Delta
=
\min_{m\neq n}
|\omega_{mn}|
>0.
\end{equation}
Here, the minimum is taken only over pairs of energy eigenstates with nonzero initial coherence, i.e. such that $\rho_{mn}(0)\neq 0$.
As shown in
Sec.~\ref{sec:protocol}, the different matrix elements $\rho_{r,mn}(t)=\bra{m}\hat\rho_r(t)\ket{n}$ evolve independently according to Eq.~\eqref{eq:general_memory_energy_basis}. Their exact time-dependent solution is given in
Eq.~\eqref{eq:general_memory_density_matrix_solution}, while their long-time
behavior follows from Eq.~\eqref{eq:protocol_density_matrix_asymptotic}.
For every pair of distinct states with nonzero initial coherence, one has
$|\omega_{mn}|\geq\Delta>0$. Therefore, all off-diagonal matrix elements
vanish algebraically at long times. In particular, when $m\neq n$, their modulus behaves as $\left|\rho_{r,mn}(t)\right|
\sim t^{-\frac{\omega_{mn}^{2}}{r^{2}+\omega_{mn}^{2}}}$, 
while their phase oscillates periodically as a function of $\log t$, with logarithmic angular frequency
$r\omega_{mn}/(r^2+\omega_{mn}^2)$. 
Since the decay exponent
$\omega_{mn}^2/(r^2+\omega_{mn}^2)$ increases monotonically with
$|\omega_{mn}|$, the slowest-decaying contribution is associated with the
smallest nonzero Bohr frequency $\Delta$. Therefore, the asymptotic approach of the full density matrix to the stationary state is controlled only by $\Delta$ and the resetting rate $r$:
\begin{equation}
\label{eq:gapped_slowest_decay}
\hat\rho_r(t)-\hat\rho_r^{\rm st}
=
O\left(
t^{-\theta}
\right),
\qquad
\theta
=
\frac{\Delta^2}{r^2+\Delta^2},
\end{equation}
 where 
$0<\theta<1$. A similar algebraic relaxation occurs for a classical diffusing particle
confined by a potential and subject to uniform memory resetting~\cite{BM2024b}.
In that case, the probability distribution approaches the Gibbs--Boltzmann
stationary state algebraically, with the slowest decay exponent $\theta_{\rm cl}=\frac{\lambda}{r+\lambda},$
where $\lambda$ is the spectral gap between the stationary mode and the first
excited mode of the Fokker--Planck operator. 
This is different from the quantum case, where the exponent $\theta$ in Eq. \eqref{eq:gapped_slowest_decay} depends on the global minimum of all spectral gaps $\Delta$.
In both the quantum and classical cases, increasing the resetting rate $r$ decreases the decay exponent $\theta$, leading to slower relaxation towards the stationary state. This is because more frequent resetting slows down the exploration of new states by repeatedly bringing the system back to previously visited ones. Interestingly, in the quantum case, the logarithmic angular frequency associated with the slowest relaxation mode, $\chi_{\rm I}(\Delta)=r\Delta/(r^2+\Delta^2)$, instead depends non-monotonically on $r$, reaching its maximum value of $1/2$ at $r=\Delta$.

The approach to the stationary state given in Eq. \eqref{eq:gapped_slowest_decay} is therefore anomalously slow
and algebraic, in sharp contrast with the exponential relaxation typically
observed under conventional resetting. 
By contrast, the diagonal matrix elements correspond to
$\omega_{nn}=0$ and therefore remain equal to their initial values at all
times, $\rho_{r,nn}(t)=\rho_{nn}(0).$
The system consequently approaches the stationary density matrix
\begin{equation}
\label{eq:gapped_nondegenerate_stationary_state}
\hat\rho_r^{\rm st}
=
\sum_n \rho_{nn}(0)\ket{n}\bra{n}.
\end{equation}
In other words, for a nondegenerate spectrum the long-time dynamics
suppresses all coherences between different energy eigenstates, while
preserving the initial populations. This can be represented schematically as
\begingroup
\setlength{\arraycolsep}{4pt}
\renewcommand{\arraystretch}{0.9}
\begin{equation}
\label{eq:gapped_density_matrix_diagonalization}
\hat{\rho}(0)
=
\begin{pmatrix}
\rho_{11}(0) & \rho_{12}(0) & \cdots & \rho_{1N}(0) \\
\rho_{21}(0) & \rho_{22}(0) & \cdots & \rho_{2N}(0) \\
\vdots & \vdots & \ddots & \vdots \\
\rho_{N1}(0) & \rho_{N2}(0) & \cdots & \rho_{NN}(0)
\end{pmatrix}
\;
\xrightarrow{\,t\to\infty\,}
\;
\hat\rho_r^{\rm st}
=
\begin{pmatrix}
\rho_{11}(0) & 0 & \cdots & 0 \\
0 & \rho_{22}(0) & \cdots & 0 \\
\vdots & \vdots & \ddots & \vdots \\
0 & 0 & \cdots & \rho_{NN}(0)
\end{pmatrix}.
\end{equation}
\endgroup
The stationary state is therefore completely determined by the initial
populations and is remarkably independent of the resetting rate $r$. In this sense, a strong memory of the initial condition is retained.
This stationary state can also be identified with the infinite-time average
of the corresponding reset-free unitary evolution. Indeed, in the energy
eigenbasis the reset-free density matrix follows from Eq. \eqref{eq:density_matrix_timet} and reads~\cite{FHS2010}
\begin{equation}
\label{eq:rho0t_gapped_systems}
\hat\rho_0(t)
=
\sum_{m,n}
\rho_{mn}(0)
\rme^{-\rmi\omega_{mn}t}
\ket{m}\bra{n}.
\end{equation}
Averaging this expression over a time interval of duration $T$ gives
\begin{align}
\frac{1}{T}
\int_0^T
\rmd t\,
\hat\rho_0(t)
=
\sum_{m,n}
\rho_{mn}(0)
\left[
\frac{1}{T}
\int_0^T
\rmd t\,
\rme^{-\rmi\omega_{mn}t}
\right]
\ket{m}\bra{n}.
\end{align}
For a nondegenerate spectrum and for large $T$, we can use the identity $\lim_{T\to\infty}
\frac{1}{T}\int_0^T\rmd t\,\rme^{-\rmi\omega_{mn}t}=\delta_{mn}$.
Hence,
\begin{equation}
\label{eq:gapped_stationary_time_average}
\hat\rho_r^{\rm st}
=
\lim_{T\to\infty}
\frac{1}{T}
\int_0^T
\rmd t\,
\hat\rho_0(t)
=
\sum_n
\rho_{nn}(0)\ket{n}\bra{n},
\end{equation}
which coincides with
Eq.~\eqref{eq:gapped_nondegenerate_stationary_state}. 
Here, the limit $T\to\infty$ means that the averaging interval is much longer than the longest relevant oscillation period, namely $T\gg 2\pi/\Delta$. In this regime, all off-diagonal contributions undergo many oscillations and average to zero.
Uniform-memory resetting therefore progressively suppresses the oscillatory
contributions associated with nonzero Bohr frequencies, while preserving the time-independent component of the reset-free dynamics.
As anticipated in Sec. \ref{sec:protocol}, this marks a crucial difference from the classical confined case \cite{BM2024b}: although the approach to the stationary state is algebraic in both settings, the classical stationary state is always the Gibbs--Boltzmann distribution, independently of the initial condition, whereas in the quantum case the stationary state is entirely determined by the initial energy populations. In this sense, the quantum case has a stronger memory.

The extension to degenerate spectra is immediate: matrix elements connecting states with the same energy have zero Bohr frequency and therefore remain equal to their initial values, while those connecting different energies decay as above. We therefore do not discuss the degenerate case further.

In the following, we illustrate these general results through two concrete systems with a discrete energy spectrum: a quantum two-level system and a quantum harmonic oscillator.

\subsection{Quantum two-level system}
\label{subsec:gapped_quantum_spin}

We first illustrate the general results derived above using the simplest nontrivial gapped quantum system: a two-level system governed by the Hamiltonian $\hat H=\Omega\hat\sigma_x$, where $\hat\sigma_x$ is the Pauli matrix along the $x$ direction. The corresponding energy eigenstates are $\ket{\uparrow_x}$ and $\ket{\downarrow_x}$, with eigenenergies $E_{\uparrow}=+\Omega$ and $E_{\downarrow}=-\Omega$, respectively.
The only nonzero Bohr frequencies are therefore
$\omega_{\uparrow\downarrow}=2\Omega$ and
$\omega_{\downarrow\uparrow}=-2\Omega$.
We consider a general initial density matrix in the energy basis,
\begin{equation}
\label{eq:gapped_spin_general_initial_density}
\hat\rho_r(0)
=
\begin{pmatrix}
\rho_{\uparrow\uparrow}(0)
&
\rho_{\uparrow\downarrow}(0)
\\
\rho_{\downarrow\uparrow}(0)
&
\rho_{\downarrow\downarrow}(0)
\end{pmatrix},
\end{equation}
where
$\rho_{\uparrow\uparrow}(0)+\rho_{\downarrow\downarrow}(0)=1$ and
$\rho_{\downarrow\uparrow}(0)=\rho_{\uparrow\downarrow}^*(0)$.
In the absence of resetting, the density matrix evolves unitarily according
to Eq.~\eqref{eq:density_matrix_timet}. Since the Hamiltonian is diagonal in the $\hat\sigma_x$ basis, one obtains~\cite{FHS2010}
\begin{equation}
\label{eq:gapped_spin_reset_free_density_matrix}
\hat\rho_0(t)
=
\begin{pmatrix}
\rho_{\uparrow\uparrow}(0)
&
\rho_{\uparrow\downarrow}(0)\rme^{-2\rmi\Omega t}
\\
\rho_{\downarrow\uparrow}(0)\rme^{2\rmi\Omega t}
&
\rho_{\downarrow\downarrow}(0)
\end{pmatrix}.
\end{equation}
Thus, the energy populations remain constant, while the coherences oscillate
indefinitely with Bohr frequencies $\pm 2\Omega$.

We now consider stochastic resetting at a rate $r$ with uniform memory.
Using the general solution given by Eq.~\eqref{eq:general_memory_factorization}, we obtain
\begin{equation}
\label{eq:gapped_spin_density_matrix}
\hat\rho_r(t)
=
\begin{pmatrix}
\rho_{\uparrow\uparrow}(0)
&
\rho_{\uparrow\downarrow}(0)f(2\Omega,t)
\\
\rho_{\downarrow\uparrow}(0)f(-2\Omega,t)
&
\rho_{\downarrow\downarrow}(0)
\end{pmatrix},
\end{equation}
where $f(\omega,t)$ is given in Eq. \eqref{eq:general_memory_kummer_solution}.
Equation~\eqref{eq:gapped_spin_density_matrix} directly illustrates the
general results derived above. 
\begin{figure}[t]
\includegraphics[width=0.48\linewidth]{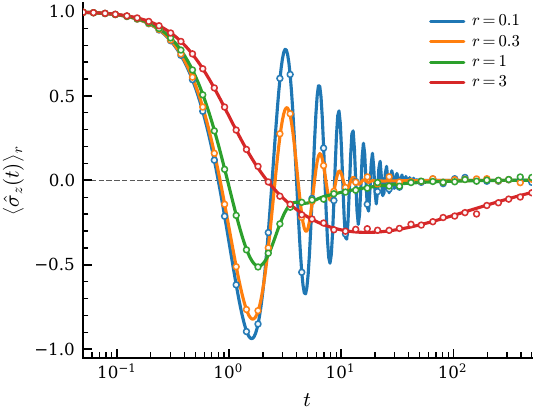}
\includegraphics[width=0.48\linewidth]{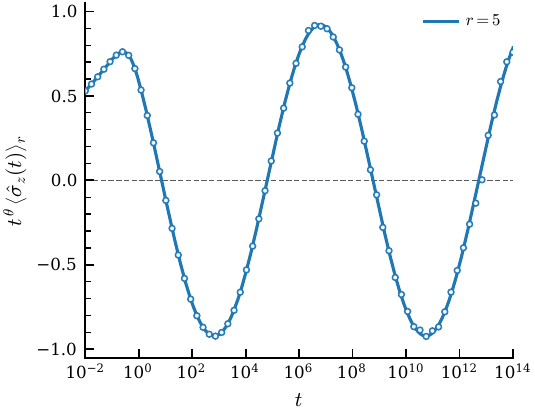}
\caption{
Reset-averaged polarization
$\langle\hat{\sigma}_z(t)\rangle_r$ for a two-level system with
$\hat H=\Omega\hat\sigma_x$, initially prepared in $\ket{\uparrow_z}$.
(Left) Exact analytical result in Eq.~\eqref{eq:sigmaz_uniform_memory}
(solid lines) and numerical simulations (circles) for different resetting
rates $r$. The polarization approaches zero through algebraically damped
oscillations.
(Right) Long-time behavior for $r=5$, after the polarization has been multiplied by $t^\theta$, with
$\theta=4\Omega^2/(r^2+4\Omega^2)$.
This rescaling removes the algebraically decaying envelope [see Eq. \eqref{eq:longtime_sigmaz_spin}] and makes the slow
oscillations, which are periodic in $\log t$, clearly visible.
In both panels, $\Omega=1$.
}
    \label{fig:uniform_twostate}
\end{figure}
The diagonal elements, corresponding to zero
Bohr frequency, remain equal to their initial values: $\rho_{r,\uparrow\uparrow}(t)=\rho_{\uparrow\uparrow}(0)$ and $\rho_{r,\downarrow\downarrow}(t)=\rho_{\downarrow\downarrow}(0)$.
By contrast, the off-diagonal elements evolve according to
Eq.~\eqref{eq:general_memory_density_matrix_solution}. In this two-level
system, the frequency gap defined in Eq.~\eqref{eq:gapped_frequency_gap} is
$\Delta=2\Omega$. Therefore, at large times, the modulus of the coherences
decays algebraically as
\begin{equation}
\left|
\rho_{r,\uparrow\downarrow}(t)
\right|
\sim
t^{-\theta},
\qquad
\theta
=
\frac{\Delta^2}{r^2+\Delta^2}
=
\frac{4\Omega^2}{r^2+4\Omega^2},
\end{equation}
while their phase oscillates periodically as a function of $\log t$, with
logarithmic angular frequency $2r\Omega/(r^2+4\Omega^2)$. Consequently, the
spin approaches the stationary density matrix predicted by
Eq.~\eqref{eq:gapped_density_matrix_diagonalization},
\begin{equation}
\label{eq:gapped_spin_stationary_state}
\hat\rho_r^{\rm st}
=
\begin{pmatrix}
\rho_{\uparrow\uparrow}(0)
&
0
\\
0
&
\rho_{\downarrow\downarrow}(0)
\end{pmatrix}.
\end{equation}
This stationary state is independent of the resetting rate $r$ and coincides with the infinite-time average of the reset-free density matrix in Eq.~\eqref{eq:gapped_spin_reset_free_density_matrix}.

To illustrate these results and enable a direct comparison with numerical
simulations, we now consider the polarization along the $z$ direction. It is obtained from
$\left\langle\hat\sigma_z(t)\right\rangle_r=
\operatorname{Tr}\left[\hat\sigma_z\hat\rho_r(t)\right].$
Since, in the $\hat\sigma_x$ basis, $\hat\sigma_z
=\ket{\uparrow_x}\bra{\downarrow_x}+\ket{\downarrow_x}\bra{\uparrow_x}$~\cite{FHS2010},
we find, for the general initial condition in
Eq.~\eqref{eq:gapped_spin_general_initial_density}, $\left\langle\hat\sigma_z(t)\right\rangle_r=2\operatorname{Re}\left[\rho_{\uparrow\downarrow}(0)f(2\Omega,t)\right].$
In Fig.~\ref{fig:uniform_twostate}, we specialize to the initial state
$\ket{\uparrow_z}$, for which
$\rho_{\uparrow\downarrow}(0)=1/2$ in the $\hat\sigma_x$ basis. Therefore,
\begin{align}
\label{eq:sigmaz_uniform_memory}
\left\langle
\hat\sigma_z(t)
\right\rangle_r
=
\operatorname{Re}
\left[
f(2\Omega,t)
\right]=
\operatorname{Re}
\left[
M\left(
\frac{2\rmi\Omega}{r+2\rmi\Omega};
1;
-\left(r+2\rmi\Omega\right)t
\right)
\right].
\end{align}
Using Eq.\eqref{eq:protocol_f_asymptotic_amplitude_phase}, we find that at large times Eq.~\eqref{eq:sigmaz_uniform_memory} behaves as
\begin{align}
\label{eq:longtime_sigmaz_spin}
\left\langle
\hat\sigma_z(t)
\right\rangle_r
\approx
C(2\Omega)
\left[
\sqrt{r^2+4\Omega^2}\,t
\right]^{
-\theta
}
\cos\left[
\frac{2r\Omega}{r^2+4\Omega^2}
\log\left(
\sqrt{r^2+4\Omega^2}\,t
\right)
+
\varphi(2\Omega)
\right],
\end{align}
where we recall $\theta=\frac{4\Omega^2}{r^2+4\Omega^2}$, $C(\omega)$ and $\varphi(\omega)$ are given in Eqs. \eqref{eq:protocol_f_amplitude} and \eqref{eq:protocol_f_phase}, respectively. 
Thus, the polarization approaches zero through oscillations (with an algebraically decaying envelope), which are periodic as a function of $\log t$.

\subsection{Quantum harmonic oscillator}
\label{subsec:gapped_quantum_harmonic_oscillator}

As a second example, we consider a quantum harmonic oscillator of mass $m$ and angular frequency $\Omega_{\rm ho}$, described by
\begin{equation}
\label{eq:gapped_qho_Hamiltonian}
\hat H=\frac{\hat p^2}{2m}+\frac{1}{2}m\Omega_{\rm ho}^2\hat x^2.
\end{equation}
Its energy eigenstates satisfy $\hat H\ket{n}=E_n\ket{n}$, with $E_n=\Omega_{\rm ho}(n+1/2)$ and Bohr frequencies $\omega_{n\ell}=(n-\ell)\Omega_{\rm ho}$. In the position representation,
\begin{equation}
\label{eq:gapped_qho_eigenfunctions}
\varphi_n(x)=\braket{x|n}=\left(\frac{m\Omega_{\rm ho}}{\pi}\right)^{1/4}\frac{H_n\left(\sqrt{m\Omega_{\rm ho}}\,x\right)}{\sqrt{2^n n!}}\exp\left(-\frac{m\Omega_{\rm ho}x^2}{2}\right),
\end{equation}
where $H_n$ is the $n$-th Hermite polynomial~\cite{FHS2010}.
We choose the normalized Gaussian initial state
\begin{equation}
\label{eq:initcond_HO}
\psi_0(x,0)=\left(\frac{2\alpha}{\pi}\right)^{1/4}\rme^{-\alpha x^2}, 
\end{equation}
and introduce the dimensionless width parameter
\begin{equation}
\eta=\frac{2\alpha}{m\Omega_{\rm ho}},
\end{equation}
where $\eta=1$ corresponds to the ground state of the harmonic oscillator. Since $\psi_0(x,0)$ is even, only even energy levels are populated, so that $\ket{\psi_0(0)}=\sum_{n=0}^{\infty}c_{2n}\ket{2n}$~\cite{FHS2010}, with
\begin{equation}
\label{eq:gapped_qho_gaussian_coefficients}
c_{2n}=\left(\frac{2\sqrt{\eta}}{1+\eta}\right)^{1/2}\frac{\sqrt{(2n)!}}{2^n n!}\left(\frac{1-\eta}{1+\eta}\right)^n, \qquad c_{2n+1}=0.
\end{equation}
For $\eta=1$, one has $c_0=1$ and $c_{2n}=0$ for all $n\geq1$.
The reset-free and reset-averaged density matrices follow directly from Eqs.~\eqref{eq:density_matrix_timet} and \eqref{eq:general_memory_factorization}:
\begin{align}
\hat\rho_0(t)&=\sum_{n,\ell=0}^{\infty}c_{2n}c_{2\ell}^*\rme^{-2\rmi(n-\ell)\Omega_{\rm ho}t}\ket{2n}\bra{2\ell},
\label{eq:gapped_qho_reset_free_density_matrix}\\
\hat\rho_r(t)&=\sum_{n,\ell=0}^{\infty}c_{2n}c_{2\ell}^*f\left(2(n-\ell)\Omega_{\rm ho},t\right)\ket{2n}\bra{2\ell},
\label{eq:gapped_qho_gaussian_density_matrix}
\end{align}
where $f(\omega,t)$ is given in Eq.~\eqref{eq:general_memory_kummer_solution}. Since only even levels are populated, the smallest nonzero Bohr frequency contributing to the dynamics is $\Delta=2\Omega_{\rm ho}$. Thus, for $\eta\neq1$, Eq.~\eqref{eq:gapped_slowest_decay} gives $\hat\rho_r(t)-\hat\rho_r^{\rm st}=O(t^{-\theta})$, with $\theta=4\Omega_{\rm ho}^2/(r^2+4\Omega_{\rm ho}^2)$, up to oscillations periodic in $\log t$ with logarithmic angular frequency $2r\Omega_{\rm ho}/(r^2+4\Omega_{\rm ho}^2)$. For the special case $\eta=1$, Eqs. \eqref{eq:gapped_qho_reset_free_density_matrix} and \eqref{eq:gapped_qho_gaussian_density_matrix} reduce, respectively, to $\hat\rho_0(t)=\ket{0}\bra{0}$ and $\hat\rho_r(t)=\ket{0}\bra{0}$ at all $t$.

At long times, all coherences between different energy levels vanish (see Eq. \eqref{eq:gapped_density_matrix_diagonalization}) and the stationary density matrix is
\begin{equation}
\label{eq:gapped_qho_stationary_state}
\hat\rho_r^{\rm st}=\sum_{n=0}^{\infty}|c_{2n}|^2\ket{2n}\bra{2n}.
\end{equation}
As predicted by the general result in Eq.~\eqref{eq:gapped_stationary_time_average}, this state is independent of $r$ and coincides with the infinite-time average of the reset-free evolution. In the position representation,
\begin{equation}
\label{eq:gapped_qho_reset_free_probability_spectral}
P_0(x,t)
=
\bra{x}\hat\rho_0(t)\ket{x}
=
\sum_{n,\ell=0}^{\infty}
c_{2n}c_{2\ell}^*
\rme^{-2\rmi(n-\ell)\Omega_{\rm ho}t}
\varphi_{2n}(x)\varphi_{2\ell}^*(x),
\end{equation}
where $P_0(x,t)$ is periodic with period $\pi/\Omega_{\rm ho}$. Averaging Eq.~\eqref{eq:gapped_qho_reset_free_probability_spectral} over one period and using $\frac{\Omega_{\rm ho}}{\pi}\int_0^{\pi/\Omega_{\rm ho}}\rmd t\,\rme^{-2\rmi(n-\ell)\Omega_{\rm ho}t}=\delta_{n\ell}$ yields
\begin{equation}
\label{eq:gapped_qho_stationary_probability_time_average}
P_r^{\rm st}(x)
=
\sum_{n=0}^{\infty}
|c_{2n}|^2
|\varphi_{2n}(x)|^2
=
\frac{\Omega_{\rm ho}}{\pi}
\int_0^{\pi/\Omega_{\rm ho}}
\rmd t\,P_0(x,t).
\end{equation}
Using Eqs.~\eqref{eq:initcond_HO} and \eqref{eq:gapped_qho_gaussian_coefficients}, Eq.~\eqref{eq:gapped_qho_reset_free_probability_spectral} can be written as $P_0(x,t)=|\psi_0(x,t)|^2$, where $\psi_0(x,t)$ is the wave function obtained by evolving the initial Gaussian state under the Hamiltonian in Eq.~\eqref{eq:gapped_qho_Hamiltonian}. This time evolution can be calculated using standard methods, for example by means of the harmonic-oscillator propagator given in Eq.~(8.1) of Ref.~\cite{FHS2010}. Performing the resulting Gaussian integral and taking the modulus squared, we obtain
\begin{equation}
\label{eq:gapped_qho_reset_free_probability}
P_0(x,t)=\sqrt{\frac{2\alpha}{\pi D_\eta(t)}}\exp\left[-\frac{2\alpha x^2}{D_\eta(t)}\right], \qquad D_\eta(t)=\cos^2(\Omega_{\rm ho}t)+\eta^2\sin^2(\Omega_{\rm ho}t).
\end{equation}
Substituting this expression into Eq.~\eqref{eq:gapped_qho_stationary_probability_time_average} gives
\begin{equation}
\label{eq:gapped_qho_stationary_probability}
P_r^{\rm st}(x)=\frac{\Omega_{\rm ho}}{\pi}\int_0^{\pi/\Omega_{\rm ho}}\rmd t\,\sqrt{\frac{2\alpha}{\pi D_\eta(t)}}\exp\left[-\frac{2\alpha x^2}{D_\eta(t)}\right].
\end{equation}

\section{Gapless Quantum Systems}
\label{sec:gapless}

We now consider quantum systems with a continuous, gapless energy spectrum. We denote
the energy eigenstates by $\ket{p}$, satisfying
\begin{equation}
\label{eq:gapless_energy_eigenstates}
\hat H\ket{p}
=
\varepsilon(p)\ket{p},
\qquad
\braket{p|p'}=\delta(p-p').
\end{equation}
The matrix elements of $\hat \rho_r(t)$ in this basis are
$\rho_r(p,p',t)=\bra{p}\hat\rho_r(t)\ket{p'}$, while the corresponding Bohr frequencies read $\omega_{pp'}=\varepsilon(p)-\varepsilon(p').$ We also define the group velocity as $v(p)=\frac{\mathrm{d}\varepsilon(p)}{\mathrm{d}p}$.
As a representative class, we consider translationally invariant systems with the power-law dispersion
\begin{equation}
\label{eq:power_law_dispersion}
\varepsilon_{\nu}(p)
=
\kappa_{\nu}|p|^{\nu},
\qquad
\kappa_{\nu}>0,
\qquad
\nu>1.
\end{equation}
The restriction $\nu>1$ ensures that the dispersion is differentiable at
$p=0$ and that the group velocity is a continuous function of momentum.

As in the gapped case, the different energy-basis matrix elements evolve
independently, and the Hamiltonian enters their dynamics only through the
corresponding Bohr frequency. Their exact time evolution is again given by Eq. \eqref{eq:general_memory_density_matrix_solution}, namely $\rho_r(p,p',t)=\rho(p,p',0)f\left(\omega_{pp'},t\right)$,
where $f(\omega,t)$ is given in
Eq.~\eqref{eq:general_memory_kummer_solution}. For every fixed pair
$(p,p')$ with $\omega_{pp'}\neq0$, it follows directly from Eq. \eqref{eq:protocol_density_matrix_asymptotic} that the corresponding matrix element vanishes algebraically at long times as
\begin{equation}
\label{eq:gapless_fixed_frequency_decay}
\left|
\rho_r(p,p',t)
\right|
\sim
t^{-\frac{\omega_{pp'}^2}{r^2+\omega_{pp'}^2}}.
\end{equation}
The crucial difference from the gapped case is that the zero-frequency sector is not separated from the nonzero Bohr frequencies by a finite gap. Instead, arbitrarily small nonzero frequencies are present, and the corresponding density-matrix elements decay increasingly slowly as $\omega\to0$ [see Eq. \eqref{eq:gapless_fixed_frequency_decay}]. Consequently, the density matrix cannot be decomposed into a time-independent contribution plus a term that vanishes uniformly at long times, as in the gapped case. Although every fixed matrix element with $\omega\neq0$ eventually decays, matrix elements associated with frequencies increasingly close to zero continue to contribute to observables, thereby preventing convergence to a stationary state.

As shown in \ref{sec:appendix_longtimebehav_gauss}, for a sufficiently
regular initial density matrix the long-time position distribution takes the
scaling form
\begin{equation}
\label{eq:gapless_position_scaling}
P_r(x,t)
\approx
\frac{r}{\log(rt)}
\mathcal V_0\left(
\frac{rx}{\log(rt)}
\right).
\end{equation}
where the function $\mathcal V_0(v)$ is given by 
\begin{equation}
\label{eq:gapless_velocity_distribution}
\mathcal V_0(v)
=
\int_{-\infty}^{\infty}
\rmd p\,
\rho(p,p,0)
\delta\left(
v-\frac{\rmd\varepsilon(p)}{\rmd p}
\right).
\end{equation}
Equation~\eqref{eq:gapless_position_scaling} shows that there is a characteristic spatial scale that
grows in time as $\log(rt)/r$, independently of the specific dispersion relation and
of the initial state. This logarithmic growth is therefore a universal
consequence of the continuous spectrum and of the uniform memory protocol.
By contrast, the shape of the spreading distribution is nonuniversal: it
depends strongly on the details of the Hamiltonian through the group velocity
$v(p)=\frac{\rmd\varepsilon(p)}{\rmd p}$ and retains a strong memory of the initial condition through $\rho(p,p,0)$. This behavior differs qualitatively from that of an
unconfined classical Brownian particle under uniform memory resetting \cite{BEM2017}, whose
long-time position distribution is universally Gaussian, independently of
the initial condition, with a width growing as $\sqrt{\log t}$. In the
quantum case, instead, the width grows as $\log t$, while the asymptotic
distribution is generally non-Gaussian, system dependent, and strongly
sensitive to the initial condition.

For the power-law dispersion in
Eq.~\eqref{eq:power_law_dispersion}, the group velocity is $v_\nu(p)= \frac{d\varepsilon_\nu(p)}{dp}=\kappa_{\nu}\nu\,\operatorname{sgn}(p)|p|^{\nu-1}.$
For $\nu>1$, this relation is monotonic and can be inverted as
\begin{equation}\label{eq:pdiv_plaw}
    p(v)=\operatorname{sgn}(v)\left(\frac{|v|}{\kappa_{\nu}\nu}\right)^{\frac{1}{\nu-1}}
\end{equation}
with Jacobian $\left|\frac{\rmd p}{\rmd v}\right|
=\frac{|v|^{\frac{2-\nu}{\nu-1}}}{(\nu-1)(\kappa_{\nu}\nu)^{\frac{1}{\nu-1}}}$.
The initial velocity distribution defined in
Eq.~\eqref{eq:gapless_velocity_distribution} therefore becomes
\begin{align}
\label{eq:gapless_power_law_velocity_distribution}
\mathcal V_0(v)
=
\frac{
|v|^{\frac{2-\nu}{\nu-1}}
}{
(\nu-1)
(\kappa_{\nu}\nu)^{\frac{1}{\nu-1}}
}\rho\big(p(v),p(v),0\big).
\end{align}
Substituting this expression into
Eq.~\eqref{eq:gapless_position_scaling}, we obtain the explicit long-time
position distribution
\begin{align}
\label{eq:gapless_power_law_position_distribution}
P_r(x,t)
\approx
\frac{r}{\log(rt)}
\frac{
\left|
\frac{rx}{\log(rt)}
\right|^{\frac{2-\nu}{\nu-1}}
}{
(\nu-1)
(\kappa_{\nu}\nu)^{\frac{1}{\nu-1}}
}
\rho\left(p\left(\frac{rx}{\log(rt)}\right),p\left(\frac{rx}{\log(rt)}\right),0\right),
\end{align}
where $p(v)$ is given in Eq. \eqref{eq:pdiv_plaw} and we recall $\rho(p,p,0)=\bra{p}\hat\rho(0)\ket{p}$ is the momentum distribution at time $t=0$.

\subsection{Quantum free particle}
\label{subsec:gapless_quantum_free_particle}

As a representative example of a gapless quantum system, we consider a quantum free
particle of mass $m$ moving on the infinite line. This corresponds to the quadratic dispersion relation $\varepsilon_2(p)=\frac{p^2}{2m}$,
obtained by setting $\nu=2$ and $\kappa_2=\frac{1}{2m}$ into Eq. \eqref{eq:power_law_dispersion}.
The associated group velocity is therefore $v_2(p)
=\frac{\rmd\varepsilon_2(p)}{\rmd p}=\frac{p}{m}.$
We again choose the normalized Gaussian initial state in Eq.~\eqref{eq:initcond_HO}, from which the initial momentum distribution follows directly by Fourier transformation: $\rho(p,p,0)=\frac{1}{\sqrt{2\pi\alpha}}\exp\left(-\frac{p^2}{2\alpha}\right)$.
Using $\nu=2$ and $\kappa_2=\frac{1}{2m}$, the initial velocity distribution in Eq. \eqref{eq:gapless_power_law_velocity_distribution} becomes Gaussian:
\begin{equation}
\label{eq:gapless_free_particle_initial_velocity_distribution}
\mathcal V_0(v)
=
m\,\rho(mv,mv,0)
=
\frac{m}{\sqrt{2\pi\alpha}}
\exp\left(
-\frac{m^2v^2}{2\alpha}
\right).
\end{equation}
Substituting this expression into
Eq.~\eqref{eq:gapless_position_scaling}, we obtain the long-time position
probability density
\begin{equation}
\label{eq:gapless_free_particle_asymptotic_probability}
P_r(x,t)
\approx
\frac{mr}{
\sqrt{2\pi\alpha}\,\log(rt)
}
\exp\left[
-\frac{
m^2r^2x^2
}{
2\alpha[\log(rt)]^2
}
\right],
\qquad
t\to\infty.
\end{equation}
The position distribution is therefore asymptotically Gaussian, with zero mean and variance $\left\langle
\hat x^2(t)\right\rangle_r\approx\frac{\alpha}{m^2r^2}[\log(rt)]^2.$
Consequently, its characteristic width grows as $\sqrt{
\left\langle\hat x^2(t)\right\rangle_r}\sim\log(rt).$

Thus, the particle does not approach a stationary position distribution but continues to spread logarithmically slowly. Uniform-memory resetting strongly suppresses the ballistic spreading of the reset-free particle, whose width grows linearly in time. As discussed above, this $\log t$ growth is faster than the $\sqrt{\log t}$ spreading found for the classical Brownian particle under uniform memory resetting \cite{BEM2017}. Moreover, while the classical asymptotic distribution is universally Gaussian, the Gaussian form obtained here results specifically from the quadratic dispersion and the Gaussian initial momentum distribution.

\section{Conclusions and Outlook}
\label{sec:conc}

In this work, we introduced a quantum stochastic resetting protocol with memory in which, at each resetting event, the system is returned to a state visited at an earlier time selected uniformly from its entire history. This construction provides a quantum counterpart of the classical preferential relocation model~\cite{BSS2014,BEM2017}, while combining coherent evolution, stochastic resetting, and temporal nonlocality. The resulting dynamics is therefore both nonunitary and non-Markovian.

Working in the energy eigenbasis, we obtained an exact solution for any density-matrix element of a general time-independent Hamiltonian and showed that the details of such Hamiltonian enter the dynamics only through the corresponding Bohr frequencies. This naturally leads to a classification into gapped and gapless quantum systems.
For \emph{gapped systems}, the relevant nonzero Bohr frequencies are bounded away from zero. Consequently, coherences between states with different energies vanish algebraically while oscillating periodically as functions of $\log t$. The slowest relaxation is controlled by the smallest relevant Bohr frequency $\Delta$, with decay exponent $\theta=\Delta^2/(r^2+\Delta^2)$. The populations in the energy basis remain equal to their initial values, as do coherences within degenerate energy eigenspaces. The stationary state is therefore obtained by projecting the initial density matrix onto the eigenspaces of the Hamiltonian. For a nondegenerate spectrum, it is simply its diagonal part in the energy basis. Remarkably, this stationary state is independent of the resetting rate $r$ and coincides with the infinite-time average of the reset-free unitary evolution. These results were illustrated concretely using a quantum two-level system and a quantum harmonic oscillator.
The dependence of the stationary state on the initial energy populations reveals a strong memory of the initial condition. This marks a crucial difference from confined classical systems under uniform memory resetting \cite{BM2024b}. Although both classical and quantum systems relax algebraically, the classical stationary state is the equilibrium Gibbs--Boltzmann distribution and is independent of the initial condition, whereas the quantum stationary state has a very strong memory of the initial condition.

For \emph{gapless systems}, the continuous spectrum contains arbitrarily small nonzero Bohr frequencies. In this case, the system does not generally reach a stationary position distribution. For translationally invariant systems, the long-time distribution universally spreads on the extremely slow scale $\log(rt)/r$. The complete large-time scaling function is instead non-universal. It depends on the specific dispersion relation and on the initial group-velocity distribution. It therefore retains a strong memory of the initial condition.
For a free quantum particle with a quadratic dispersion relation and a Gaussian initial state, the long-time position distribution is Gaussian, with a standard deviation growing as $\log t$, rather than linearly in $t$ as in the absence of resetting. Similarly, for a classical unconfined Brownian particle, resetting with memory changes the usual $\sqrt{t}$ growth of the width to $\sqrt{\log t}$, with an asymptotically Gaussian distribution independently of the initial condition~\cite{BEM2017}. Thus, in both cases, resetting with memory effectively replaces $t$ by $\log t$ in the long-time spreading behavior.

The spectral distinction therefore determines the qualitative long-time behavior in the cases studied here: a finite minimum nonzero Bohr frequency leads to algebraic relaxation towards a stationary state, whereas a continuous spectrum can lead to persistent logarithmic spreading. 
In both cases, however, the repeated sampling of an expanding history strongly suppresses the underlying quantum dynamics. This contrasts with conventional quantum resetting to a fixed state~\cite{MSM2018,RTLG2018}, which has a renewal structure and typically produces an exponentially fast approach to an $r$-dependent nonequilibrium stationary state.

More complicated spectra may exhibit a combination of gapped and gapless behaviors, for instance when discrete and continuous energy levels coexist, or when a discrete spectrum contains arbitrarily small Bohr frequencies, as in the hydrogen atom. In these cases, the discrete part of the spectrum gives rise to a nonzero stationary contribution, provided it is populated by the initial state. Very small Bohr frequencies or the continuous part of the spectrum may instead give rise to an additional time-dependent contribution of the form discussed in Eq.~\eqref{eq:gapless_position_scaling}, which, however, vanishes as $t\to\infty$, albeit extremely slowly.

Several interesting extensions follow naturally from the present work. It would be interesting to investigate nonuniform memory kernels that favor either recent or remote portions of the history (such as those proposed in the classical setting in Ref. \cite{BEM2017}). A particularly interesting direction is to replace the uniform kernel by a general normalized memory kernel,
$K(\tau,t)=\frac{\phi(\tau)}{\int_0^t \rmd s\,\phi(s)}$ with 
$\int_0^t \rmd\tau\,K(\tau,t)=1$ in Eq. \eqref{eq:uniform_memory}. The last term in Eq. \eqref{eq:uniform_memory} generalizes to
$r\int_0^t \rmd\tau\,K(\tau,t)\hat\rho_r(\tau)$. 
Depending on the choice of $\phi(\tau)$, the resetting protocol can preferentially sample either remote or recent portions of the history. In the classical diffusive problem, Ref.~\cite{BEM2017} showed that this temporal bias produces a remarkably broad range of long-time behaviors. It would be interesting to determine how this hierarchy is modified in the quantum setting.

It will be interesting to study interacting many-body systems, where memory may affect correlations and entanglement. 
In particular, the fact that resetting with memory slows down spreading without stopping it, unlike conventional resetting, which typically arrests it, suggests possible applications to the control of quantum transport. Further directions include time-dependent Hamiltonians, open quantum systems, and state- or observable-dependent memory kernels. Quantum resetting with memory thus provides a simple framework in which coherent evolution, stochastic control, and temporal nonlocality coexist, offering a route to generate ultraslow dynamics while preserving information about the initial quantum state.

\section*{Acknowledgments}
We acknowledge support from ANR Grant No. ANR-23-CE30-0020-01 EDIPS. M.K. acknowledges support from the Department of Atomic Energy, Government of India, under Project No. RTI4001. M. K. thanks the hospitality of Laboratoire de Physique Théorique et Modèles Statistiques (LPTMS), University Paris-Saclay and Collège de France, PSL Research University where a major part of the work took place.

\appendix
\addtocontents{toc}{\protect\appendixtocformat}

\section{Equivalence between the renewal equation and the ``Master equation" }\label{app:equiv}

This appendix relates the renewal description of quantum resetting in Ref.~\cite{MSM2018} to a ``Master equation" \cite{RTLG2018}, i.e., we derive the equivalence between Eq.~\eqref{eq:renewal_standard_resetting} and Eq.~\eqref{eq:master_standard_resetting} of the main text. We write $\hat\rho_0(t)$ for the density matrix evolving \emph{without} resetting, and $\hat\rho_r(t)$ for the density matrix averaged over the resetting process.
We start from the renewal equation, given in Eq.~\eqref{eq:renewal_standard_resetting}. Differentiating it with respect to time, we obtain
\begin{equation}
\label{eq:renewal-derivative}
\begin{split}
\frac{\mathrm{d}\hat\rho_r(t)}{\mathrm{d}t}
&=-r\rme^{-rt}\hat\rho_0(t)
+\rme^{-rt}\frac{\mathrm{d}\hat\rho_0(t)}{\mathrm{d}t}
+r\rme^{-rt}\hat\rho_0(t)=\rme^{-rt}\frac{\mathrm{d}\hat\rho_0(t)}{\mathrm{d}t}.
\end{split}
\end{equation}
The reset-free density matrix obeys the von Neumann equation
\begin{equation}
\label{eq:von-neumann}
\frac{\mathrm{d}\hat\rho_0(t)}{\mathrm{d}t}
=-\rmi\comm{\hat H}{\hat\rho_0(t)}.
\end{equation}
Substituting Eq.~\eqref{eq:von-neumann} into Eq.~\eqref{eq:renewal-derivative}, we find
\begin{equation}
\label{eq:renewal-substitution}
\frac{\mathrm{d}\hat\rho_r(t)}{\mathrm{d}t}
=-\rmi\rme^{-rt}\comm{\hat H}{\hat\rho_0(t)}
=-\rmi\comm{\hat H}{\rme^{-rt}\hat\rho_0(t)}.
\end{equation}
From the renewal equation in Eq.~\eqref{eq:renewal_standard_resetting} we have
\begin{equation}
\rme^{-rt}\hat\rho_0(t)
=\hat\rho_r(t)-r\int_0^t\mathrm{d}\tau\,\rme^{-r\tau}\hat\rho_0(\tau).
\end{equation}
Substituting this expression into Eq.~\eqref{eq:renewal-substitution}, and using the linearity of the commutator, we obtain
\begin{equation}
\label{eq:renewal-commutator}
\begin{split}
\frac{\mathrm{d}\hat\rho_r(t)}{\mathrm{d}t}
&=-\rmi\comm{\hat H}{\hat\rho_r(t)}
+\rmi r\comm{\hat H}{\int_0^t\mathrm{d}\tau\,\rme^{-r\tau}\hat\rho_0(\tau)}\\
&=-\rmi\comm{\hat H}{\hat\rho_r(t)}
+\rmi r\int_0^t\mathrm{d}\tau\,\rme^{-r\tau}\comm{\hat H}{\hat\rho_0(\tau)}\\
&=-\rmi\comm{\hat H}{\hat\rho_r(t)}
-r\int_0^t\mathrm{d}\tau\,\rme^{-r\tau}\frac{\mathrm{d}\hat\rho_0(\tau)}{\mathrm{d}\tau},
\end{split}
\end{equation}
where in the last step we used again the von Neumann equation, Eq.~\eqref{eq:von-neumann}.
The integral in Eq.~\eqref{eq:renewal-commutator} can be evaluated by integration by parts and gives
\begin{equation}
\label{eq:renewal-integration-parts}
\begin{split}
-r\int_0^t\mathrm{d}\tau\,\rme^{-r\tau}
\frac{\mathrm{d}\hat\rho_0(\tau)}{\mathrm{d}\tau}
=-r\left[\rme^{-rt}\hat\rho_0(t)
+r\int_0^t\mathrm{d}\tau\,\rme^{-r\tau}\hat\rho_0(\tau)\right]
+r\hat\rho(0)=-r\hat\rho_r(t)+r\hat\rho(0),
\end{split}
\end{equation}
where we used Eq.~\eqref{eq:renewal_standard_resetting} in the last step, and $\hat\rho(0)$ denotes the initial density matrix to which the system is returned at each resetting event.
Substituting Eq.~\eqref{eq:renewal-integration-parts} into Eq.~\eqref{eq:renewal-commutator}, we finally obtain
\begin{equation}
\label{eq:reset-master}
\frac{\mathrm{d}\hat\rho_r(t)}{\mathrm{d}t}
=-\rmi\comm{\hat H}{\hat\rho_r(t)}
-r\hat\rho_r(t)
+r\hat\rho(0),
\end{equation}
which coincides with Eq.~\eqref{eq:master_standard_resetting} of the main text.

\subsection{Lindblad form of classical resetting}\label{subsec:lindblad}

We now show that Eq.~\eqref{eq:reset-master} has Lindblad form~\cite{GKSL1976, Lindblad1976, BreuerPetruccione2006, Carmichael2002}. Let $\{\ket{\phi_\mu}\}$ be a complete orthonormal basis and define the jump operators
\begin{equation}
\hat L_\mu=\sqrt{r}\,\ket{\psi_0(0)}\bra{\phi_\mu}\,,
  \label{eq:jump-operator}
\end{equation}
where we recall that $\ket{\psi_0(0)}$ is the initial state that the system is also reset to, i.e., 
$\hat{\rho}_0(0) =\ket{\psi_0(0)} \bra{\psi_0(0)}$.
The corresponding Lindblad equation is
\begin{equation}
  \frac{\mathrm{d}\hat \rho_r}{\mathrm{d}t}
  =-\mathrm{i}\comm{\hat H}{\hat \rho_r}
  +\sum_\mu \hat L_\mu\hat \rho_r \hat L_\mu^\dagger
  -\frac{1}{2}\sum_\mu
   \acomm{\hat L_\mu^\dagger \hat L_\mu}{\hat \rho_r}.
  \label{eq:lindblad}
\end{equation}
The first dissipative term in Eq. \eqref{eq:lindblad} is
\begin{align}
  \sum_\mu \hat L_\mu\hat \rho_r \hat L_\mu^\dagger
  &=r\sum_\mu
    \ket{\psi_0(0)}\bra{\phi_\mu}
    \hat \rho_r \ket{\phi_\mu}\bra{\psi_0(0)}  =r\proj{\psi_0(0)}\Tr(\hat \rho_r)\nonumber\\
   &=r\proj{\psi_0(0)},
   \label{eq:lin1}
\end{align}
where $\Tr(\hat \rho_r)=1$ has been used in Eq \eqref{eq:lin1}. Also,
\begin{align}
\label{eq:Lmu}
  \sum_\mu \hat L_\mu^\dagger \hat L_\mu  &=r\sum_\mu\ket{\phi_\mu}\bra{\phi_\mu}
   =r\mathbb{\hat I}.
\end{align}
Consequently, using Eq \eqref{eq:Lmu}, we get
\begin{equation}
\label{eq:lin2}
  -\frac{1}{2}\sum_\mu
  \acomm{\hat L_\mu^\dagger \hat L_\mu}{\rho_r}
  =-\frac{r}{2}\acomm{\mathbb{\hat I}}{\hat\rho_r}
  =-r\hat\rho_r.
\end{equation}
Substituting Eqs \eqref{eq:lin1} and \eqref{eq:lin2} into Eq.~\eqref{eq:lindblad} reproduces Eq.~\eqref{eq:reset-master}. Thus, the renewal equation \cite{MSM2018} and the Lindblad equation describe the same Poissonian quantum-resetting dynamics \cite{RTLG2018}.

\section{Derivation of the master equation for uniform-memory resetting}
\label{app:uniform_memory_derivation}

In this appendix, we derive Eq.~\eqref{eq:uniform_memory} of the main text directly from the stochastic resetting protocol with uniform memory introduced in Sec.~\ref{sec:protocol}. In particular, we show that the density matrix averaged over all realizations of the resetting process satisfies Eq.~\eqref{eq:uniform_memory}.

Let $\hat\rho_0(q)$ denote the density matrix obtained by evolving the initial state $\hat\rho(0)$ unitarily for a time $q$, in the absence of resetting. It satisfies the von Neumann equation
\begin{equation}
\frac{\mathrm{d} \hat\rho_0(q)}{\mathrm{d} q}
=-\rmi[\hat H,\hat\rho_0(q)].
\label{eq:rho0_von_neumann}
\end{equation}
For a given realization of the resetting process, the system always remains on the same unitary trajectory generated by $\hat H$ from the initial state. Indeed, between resetting events the system evolves under the same Hamiltonian $\hat H$, while each reset simply brings it back to a state previously visited along this trajectory. The subsequent unitary evolution therefore continues along the same trajectory, rather than generating a new one. Consequently, the state can always be written as $\hat\rho_0(q(t))$, where $q(t)$ is a stochastic effective time. Between resetting events, $q(t)$ increases at unit rate, while a reset occurring at time $t$ to a uniformly chosen past time $\tau\in[0,t]$ corresponds to setting $q(t^+)=q(\tau)$.

The reset-averaged density matrix is defined as
\begin{equation}
    \hat\rho_r(t)=\left\langle \hat\rho_0(q(t)) \right\rangle,
    \label{eq:rho_reset_average}
\end{equation}
where the average is over all realizations of the resetting process.
We now consider the evolution during a short time interval $\mathrm{d} t$. With probability $1-r\,\mathrm{d} t$, no reset occurs and $q(t+\mathrm{d} t)=q(t)+\mathrm{d} t$. With probability $r\,\mathrm{d} t$, a reset occurs and a time $\tau$ is chosen uniformly in $[0,t]$, so that the state becomes $\hat\rho_0(q(\tau))$. Therefore we can write
\begin{align}
    \hat\rho_r(t+\mathrm{d} t)
    &=
    (1-r\,\mathrm{d} t)
    \left\langle
    \hat\rho_0(q(t)+\mathrm{d} t)
    \right\rangle
    +\frac{r\,\mathrm{d} t}{t}
    \int_0^t\mathrm{d}\tau\,
    \left\langle
    \hat\rho_0(q(\tau))
    \right\rangle
    +O(\mathrm{d} t^2),
    \label{eq:rho_short_time}
\end{align}
where the factor $t^{-1}\int_0^t \mathrm{d}\tau$ represents the uniform average over all past times $\tau\in[0,t]$.
Expanding the first term on the right-hand side of Eq.~\eqref{eq:rho_short_time} to first order in $\mathrm{d} t$ and using Eq.~\eqref{eq:rho0_von_neumann} we obtain
\begin{align}
    \left\langle
    \hat\rho_0(q(t)+\mathrm{d} t)
    \right\rangle\approx
    \hat\rho_r(t)
    +
    \mathrm{d} t
    \left\langle
    \frac{\partial \hat\rho_0(q)}{\partial q}
    \bigg|_{q=q(t)}
    \right\rangle=
    \hat\rho_r(t)
    -i\,\mathrm{d} t\,
    [\hat H,\hat\rho_r(t)].
\end{align}
Moreover, by definition, $\langle \hat\rho_0(q(\tau))\rangle=\hat\rho_r(\tau)$. Substituting these results into Eq.~\eqref{eq:rho_short_time}, subtracting $\hat\rho_r(t)$ and taking the limit $\mathrm{d} t\to0$, we obtain
\begin{equation}
    \frac{\mathrm{d} \hat\rho_r(t)}{\mathrm{d} t}
    =
    -i[\hat H,\hat\rho_r(t)]
    -r\hat\rho_r(t)
    +\frac{r}{t}\int_0^t\mathrm{d}\tau\,
    \hat\rho_r(\tau),
    \label{eq:memory_master}
\end{equation}
which coincides with Eq.~\eqref{eq:uniform_memory} of the main text.

\section{Event-driven Monte Carlo simulation of the uniform memory protocol}
\label{app:uniform_memory_numerics}

In this appendix, we describe the event-driven Monte Carlo procedure used to verify the analytical result in Eq.~\eqref{eq:sigmaz_uniform_memory} for the quantum two-level system under Poissonian resetting with uniform memory. We denote the density matrix along the $k$-th realization of the dynamics by $\hat\rho_r^{(k)}(t)$. Each realization is specified by a sequence of resetting times $\{T_1,T_2,\ldots\}$ and selected past times $\{\tau_1,\tau_2,\ldots\}$, where $\tau_j$ is sampled uniformly in $[0,T_j]$ at the $j$-th reset. The density matrix considered in the analytical treatment is the ensemble average $\hat\rho_r(t)=\mathbb{E}\left[
\hat\rho_r^{(k)}(t)\right]$,
where the expectation is taken over both the Poissonian resetting times and the past times selected at each event.

For the two-level system in Sec.~\ref{subsec:gapped_quantum_spin}, the Hamiltonian is $\hat H=\Omega\hat\sigma_x$. In the absence of resetting, the density matrix evolves along the unitary orbit according to Eq. \eqref{eq:gapped_spin_reset_free_density_matrix}, namely
\begin{equation}
\label{eq:numerical_reset_free_orbit}
\hat\rho_0(s)
=
\rme^{-\rmi\hat Hs}
\hat{\rho}(0)
\rme^{\rmi\hat Hs}
=
\begin{pmatrix}
\rho_{\uparrow\uparrow}(0)
&
\rho_{\uparrow\downarrow}(0)\rme^{-2\rmi\Omega s}
\\
\rho_{\downarrow\uparrow}(0)\rme^{2\rmi\Omega s}
&
\rho_{\downarrow\downarrow}(0)
\end{pmatrix},
\end{equation}
where $s$ is the time elapsed along the reset-free evolution. 

We now consider a single stochastic realization of the resetting process with memory. In this case, the state no longer follows the reset-free evolution in Eq.~\eqref{eq:numerical_reset_free_orbit} as a function of the physical time $t$. However, by construction, each reset returns the system to a state that was previously occupied and that therefore belongs to the same unitary orbit generated from the initial condition. Following what is shown rigorously in Ref.~\cite{MUB2019}, we may thus introduce an effective time $q^{(k)}(t)$ such that the state at time $t$ along the $k$-th realization can always be written as 
\begin{equation}
\label{eq:effective_time_definition}
\hat\rho_r^{(k)}(t)=\hat\rho_0\left(q^{(k)}(t)\right).
\end{equation}
The quantity $q^{(k)}(t)$ therefore identifies the point of the reset-free unitary orbit occupied by the system at physical time $t$ and should not be interpreted as an additional physical time. In the following, we consider a fixed realization and suppress the index $k$.

Let $T_1<T_2<\cdots$ be the resetting times, with $T_0=0$, and define $\bar q_j=q(T_j^+)$ as the effective time immediately after the $j$-th reset, with $\bar q_0=0$. Between two consecutive resetting events, the effective time increases linearly. In particular, if the system is at effective time $\bar q_j$ immediately after the reset at $T_j$, then $q(t)=\bar q_j+t-T_j$ for $T_j\leq t<T_{j+1}$. This follows directly from the property of the unitary evolution, since for $s=t-T_j$ one has $\hat\rho_r(T_j+s)
=\rme^{-\rmi\hat Hs}\hat\rho_0(\bar q_j)\rme^{\rmi\hat Hs}=\hat\rho_0(\bar q_j+s).$
At the $j$-th resetting event, a past physical time $\tau_j$ is sampled uniformly in $[0,T_j]$. To determine the state occupied at that time, we identify the interval containing $\tau_j$. Let $\ell$ satisfy $T_\ell\leq\tau_j<T_{\ell+1}$, with $0\leq\ell\leq j-1$. Since the system started from the effective time $\bar q_\ell$ immediately after the reset at $T_\ell$ and subsequently evolved unitarily for a duration $\tau_j-T_\ell$, the effective time immediately after the $j$-th reset is
\begin{equation}
\label{eq:general_effective_time_update}
\bar q_j
=
q(T_j^+)
=
q(\tau_j)
=
\bar q_\ell+\tau_j-T_\ell.
\end{equation}
For the first reset, $\ell=0$, and Eq.~\eqref{eq:general_effective_time_update} simply gives $\bar q_1=\tau_1$. For later resets, one generally has $q(\tau_j)\neq\tau_j$, because the state occupied at the selected physical time may already contain the effects of previous resetting events. All resets preceding $T_\ell$ are encoded in $\bar q_\ell$. Therefore, the complete memory of a realization is contained in the stored pairs $\{T_j,\bar q_j\}$ for all $j$.

The resetting times are generated by a Poisson process of rate $r$. Hence, the inter reset intervals $\Delta T_j=T_j-T_{j-1}$ are independent random variables distributed according to $R(\Delta T)=r\rme^{-r\Delta T}$. For each realization, the event-driven algorithm proceeds as follows:

\begin{enumerate}

\item Set $T_0=0$ and $\bar q_0=0$.

\item Draw an inter reset interval $\Delta T_j$ from
$R(\Delta T)=r\rme^{-r\Delta T}$ and set
$T_j=T_{j-1}+\Delta T_j$.

\item Evolve the system unitarily from $T_{j-1}^+$ to $T_j^-$. Since the effective time immediately after the previous reset is $\bar q_{j-1}$, one has $q(T_j^-)=\bar q_{j-1}+\Delta T_j,$
and therefore $\hat\rho_r(T_j^-)
=\rme^{-\rmi\hat H\Delta T_j}\hat\rho_r(T_{j-1}^+)
\rme^{\rmi\hat H\Delta T_j}=\hat\rho_0\left(\bar q_{j-1}+\Delta T_j\right)$, with $\hat\rho_0(s)$ given in Eq. \eqref{eq:numerical_reset_free_orbit}.

\item Draw a past time $\tau_j$ uniformly in $[0,T_j]$ and find the index $\ell$ such that
$T_\ell\leq\tau_j<T_{\ell+1}$.

\item Update the effective time according to $\bar q_j=\bar q_\ell+\tau_j-T_\ell,$
and reset the system to $\hat\rho_r(T_j^+)=\hat\rho_0(\bar q_j)$.

\item Store the pair $(T_j,\bar q_j)$ and repeat the procedure until the final observation time is reached.

\end{enumerate}
Once the reset history has been generated, the trajectory is evaluated at a prescribed set of observation times $t_n$. For each $t_n$, the algorithm finds the most recent reset time $T_j$ such that $T_j\leq t_n<T_{j+1}$ and reconstructs the effective time as $q(t_n)=\bar q_j+t_n-T_j$. The density matrix is then obtained directly from the reset-free solution as $\hat\rho_r^{(k)}(t_n)=\hat\rho_0(q(t_n))$. Finally, the observable of interest is evaluated for each realization and averaged over the ensemble. The resulting numerical averages are shown as dots in the left panel of Fig.~\ref{fig:uniform_twostate}.

\subsection{Backward Monte Carlo sampling at long times}
\label{app:backward_sampling}

The forward event-driven algorithm described above requires generating the entire resetting history up to time $t$, involving on average $O(rt)$ resetting events. To reach much longer times, as in the right panel of Fig.~\ref{fig:uniform_twostate}, we instead use an equivalent backward sampling procedure based on an exact recursion relation for the effective time $q(t)$.

As discussed above, the system always remains on the same unitary orbit, so that a given realization $k$ at time $t$ can be written as $\hat\rho_r^{(k)}(t)=\hat\rho_0(q^{(k)}(t))$. We can therefore sample the state at a given observation time $t$ by sampling the random variable $q(t)$ directly, without constructing the entire resetting history. In the following, we suppress the realization index $k$.

\begin{figure}[t]
    \centering
    \includegraphics[width=0.45\columnwidth]{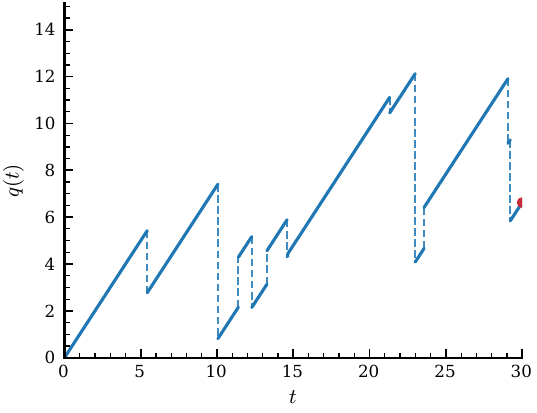}
    \hfill
    \includegraphics[width=0.45\columnwidth]{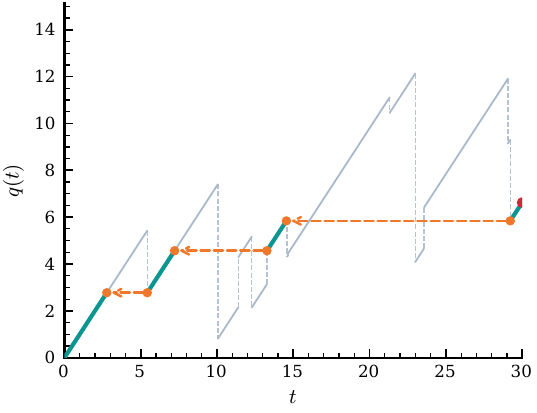}
    \caption{
    Illustration of the forward and backward sampling procedures for a single realization of uniform-memory resetting, with resetting rate $r=0.7$ and observation time $T=30$.
    (Left) Forward trajectory of the effective time $q(t)$, which increases linearly between resetting events and jumps to previously visited values at each reset.
    (Right) Backward reconstruction of the same trajectory, starting from the final observation time $T$. The green segments represent the ancestral portions of the trajectory contributing to $q(T)$, while the orange dashed arrows indicate the backward jumps to the past times selected at the corresponding resetting events. The remaining parts of the forward trajectory are shown in gray.
    The effective time $q(T)$ is obtained by summing the durations of the green segments, without reconstructing the full resetting history.}
    \label{fig:genealogy}
\end{figure}

Consider the last resetting event before the observation time $t$. To sample its occurrence, we draw an auxiliary random variable $\Delta$ from the exponential distribution with density $r\mathrm{e}^{-r\Delta}$. If $\Delta\geq t$, this represents the absence of resetting up to time $t$, which has probability $\mathrm{e}^{-rt}$, and $q(t)=t$. Otherwise, the last reset occurred at time $t-\Delta$ and returned the system to a state occupied at a past time $\tau$, chosen uniformly in $[0,t-\Delta]$. The subsequent unitary evolution lasts for a time $\Delta$, giving $q(t)=\Delta+q(\tau)$. By the independent increments of the Poisson process and the independent selection of the past time, conditioning on this last reset does not change the distribution of the history up to $\tau$. Hence, $q(\tau)$ can be sampled according to the same rule at the earlier observation time $\tau$. We therefore obtain the exact recursion relation
\begin{equation}
\label{eq:backward_recursion}
q(t)\overset{d}{=}
\begin{cases}
t, & \Delta\geq t,\\
\Delta+q(\tau), & \Delta<t,
\end{cases}
\qquad
\Delta\sim\mathrm{Exp}(r), \quad
\tau\mid\Delta\sim\mathrm{Uniform}(0,t-\Delta)
\quad\text{if }\Delta<t,
\end{equation}
where $\overset{d}{=}$ denotes equality in distribution, $\sim$ indicates that a random variable is drawn from the specified distribution, and $q(\tau)$ on the right-hand side is sampled recursively according to the same rule.
This relation provides a simple backward algorithm: starting from the observation time $t$, we draw $\Delta$ and, if $\Delta<t$, sample $\tau$ and repeat the procedure at the earlier time $\tau$, using fresh independent random draws at each step. The recursion terminates when the sampled interval $\Delta$ is greater than or equal to the current backward time. The value of $q(t)$ is obtained by adding the intervals $\Delta$ from all nonterminal steps and the remaining backward time at the terminal step. An example of a forward trajectory and its corresponding backward reconstruction is shown in Fig.~\ref{fig:genealogy}.

The advantage of this backward construction is its much lower computational cost. In a forward simulation, the average number of resetting events up to time $t$ is $O(rt)$. In the backward procedure, instead, when $rt\gg1$, the typical sampled interval $\Delta\sim1/r$ is much smaller than $t$. Therefore, the last reset occurred at time $t-\Delta\simeq t$, and the uniformly sampled past time can be written as $\tau=U(t-\Delta)\simeq Ut$, where $U$ is a random variable uniformly distributed in $[0,1]$. Thus, each backward step reduces the current time by a random multiplicative factor. After $n$ steps, we have approximately $t_n\simeq t\prod_{j=1}^n U_j$, and hence $\log(rt_n)\simeq\log(rt)+\sum_{j=1}^n\log U_j$. Since $\langle\log U\rangle=-1$, we obtain $\langle\log(rt_n)\rangle\simeq\log(rt)-n$. Once the current time becomes of order $1/r$, corresponding to $\log(rt_n)=O(1)$, the probability of termination at each step is of order one. Therefore, the typical number of backward steps scales as $n=O(\log(rt))$, compared with $O(rt)$ events in the forward simulation.

\section{Long-time position distribution for a generic dispersion relation}
\label{sec:appendix_longtimebehav_gauss}

In this Appendix, we derive the long-time position distribution of a
translationally invariant quantum particle with a generic dispersion
relation $\varepsilon(p)$ and an arbitrary initial state $\hat \rho_0(0)$. We consider a Hamiltonian of the
form introduced in Eq.~\eqref{eq:gapless_energy_eigenstates}, whose momentum
eigenstates satisfy $\hat H\ket{p}=\varepsilon(p)\ket{p}$ and
$\braket{p|p'}=\delta(p-p')$. We assume that $\varepsilon(p)$ is
differentiable in the region of momentum space explored by the initial
state.
We denote the momentum-space matrix elements of the reset-averaged density
matrix by $\rho_r(p,p',t)=\bra{p}\hat\rho_r(t)\ket{p'}$ and those of the
initial density matrix by
$\rho(p,p',0)=\bra{p}\hat\rho(0)\ket{p'}$. Using the general
factorization in Eq.~\eqref{eq:general_memory_factorization}, their time
evolution is
\begin{equation}
\label{eq:generic_dispersion_density_matrix}
\rho_r(p,p',t)
=
\rho(p,p',0)
f\left(
\varepsilon(p)-\varepsilon(p'),t
\right),
\end{equation}
where the corresponding Bohr frequency is
$\omega_{pp'}=\varepsilon(p)-\varepsilon(p')$ and $f(\omega,t)$ is given in
Eq.~\eqref{eq:general_memory_kummer_solution}.

We first express the position probability density
$P_r(x,t)=\bra{x}\hat\rho_r(t)\ket{x}$ in terms of the momentum-space density
matrix. Using $\braket{x|p}=(2\pi)^{-1/2}\mathrm{e}^{\mathrm{i}px}$, we
obtain
\begin{equation}
\label{eq:initialPr_APP}
P_r(x,t)
=
\frac{1}{2\pi}
\int_{-\infty}^{\infty}
\mathrm{d}p
\int_{-\infty}^{\infty}
\mathrm{d}p'\,
\mathrm{e}^{\mathrm{i}(p-p')x}
\rho_r(p,p',t).
\end{equation}
We now introduce the Fourier transform
$\widetilde P_r(q,t)=\int_{-\infty}^{\infty}\mathrm{d}x\,
\mathrm{e}^{-\mathrm{i}qx}P_r(x,t)$, with inverse transform
$P_r(x,t)=\frac{1}{2\pi}\int_{-\infty}^{\infty}\mathrm{d}q\,
\mathrm{e}^{\mathrm{i}qx}\widetilde P_r(q,t)$. Taking the Fourier transform
of Eq.~\eqref{eq:initialPr_APP} and using the identity
$\delta(k)=\frac{1}{2\pi}\int_{-\infty}^{\infty}\mathrm{d}x\,
\mathrm{e}^{\mathrm{i}kx}$, we find
\begin{equation}
\label{eq:generic_characteristic_delta}
\widetilde P_r(q,t)
=
\int_{-\infty}^{\infty}
\mathrm{d}p
\int_{-\infty}^{\infty}
\mathrm{d}p'\,
\rho_r(p,p',t)
\delta(p-p'-q).
\end{equation}
To evaluate the delta function, we introduce the central and relative
momenta $P=(p+p')/2$ and $Q=p-p'$. The inverse relations are
$p=P+Q/2$ and $p'=P-Q/2$. Performing the change of variables
$(p,p')\to(P,Q)$, whose Jacobian has absolute value one,
Eq.~\eqref{eq:generic_characteristic_delta} becomes
\begin{align}
\widetilde P_r(q,t)
=
\int_{-\infty}^{\infty}
\mathrm{d}P
\int_{-\infty}^{\infty}
\mathrm{d}Q\,
\rho_r\left(
P+\frac{Q}{2},
P-\frac{Q}{2},
t
\right)
\delta(Q-q).
\end{align}
Performing the integral over $Q$, renaming the remaining integration
variable $P$ as $p$, and using
Eq.~\eqref{eq:generic_dispersion_density_matrix}, we obtain
\begin{equation}
\label{eq:generic_initial_characteristic_exact}
\widetilde P_r(q,t)
=
\int_{-\infty}^{\infty}
\mathrm{d}p\,
\rho\left(
p+\frac{q}{2},
p-\frac{q}{2},
0
\right)
f\left(
\varepsilon\left(p+\frac{q}{2}\right)
-
\varepsilon\left(p-\frac{q}{2}\right),
t
\right).
\end{equation}
Equation~\eqref{eq:generic_initial_characteristic_exact} is exact at all times. We now
analyze its long-time behavior. Since the large-scale behavior of the
position distribution is controlled by the small-$q$ behavior of its
Fourier transform, we expand the energy difference appearing in the first
argument of $f$ for small $q$. This gives
\begin{equation}
\label{eq:generic_dispersion_group_velocity_expansion}
\varepsilon\left(p+\frac{q}{2}\right)
-
\varepsilon\left(p-\frac{q}{2}\right)
\approx
qv(p),
\qquad
v(p)=\frac{\mathrm{d}\varepsilon(p)}{\mathrm{d}p}.
\end{equation}
Here $v(p)$ is the group velocity associated with momentum $p$.

We next determine the large-time behavior of $f(\omega,t)$. From
Eq.~\eqref{eq:protocol_f_asymptotic}, the leading asymptotic form of the
Kummer function is
\begin{equation}
\label{eq:largetimef_APP}
f(\omega,t)
\approx
\frac{
\left[
\left(r+\mathrm{i}\omega\right)t
\right]^{-\chi(\omega)}
}{
\Gamma\left(1-\chi(\omega)\right)
}
=
\exp\left\{
-\chi(\omega)\log\left[\left(r+\mathrm{i}\omega\right)t\right]
-\log\Gamma\left(1-\chi(\omega)\right)
\right\},
\end{equation}
where we recall $\chi(\omega)=\frac{\mathrm{i}\omega}{r+\mathrm{i}\omega}$. At long
times, the dominant contribution comes from increasingly small frequencies.
We therefore consider the joint limit of large $t$ and small $\omega$ while
keeping $\omega\log(rt)$ fixed. Since
$\chi(\omega)\approx\mathrm{i}\omega/r$,
$\log[(r+\mathrm{i}\omega)t]\approx\log(rt)$, and
$\Gamma(1-\chi(\omega))\approx1$ in this limit, the leading contribution is
\begin{equation}
\label{eq:generic_dispersion_kummer_scaling}
f(\omega,t)
\approx
\exp\left[
-\frac{\mathrm{i}\omega}{r}\log(rt)
\right]
=
\mathrm{e}^{-\mathrm{i}\omega t_{\mathrm{eff}}(t)},
\qquad
t_{\mathrm{eff}}(t)
=
\frac{\log(rt)}{r}.
\end{equation}
Combining Eqs.~\eqref{eq:generic_dispersion_group_velocity_expansion} and
\eqref{eq:generic_dispersion_kummer_scaling}, we find
\begin{equation}
\label{eq:generic_initial_kummer_scaling}
f\left(
\varepsilon\left(p+\frac{q}{2}\right)
-
\varepsilon\left(p-\frac{q}{2}\right),
t
\right)
\approx
\mathrm{e}^{-\mathrm{i}q t_{\mathrm{eff}}(t)v(p)}.
\end{equation}
Moreover, since the relevant values of $q$ vanish at long times, we may use
$\rho\left(p+\frac{q}{2},p-\frac{q}{2},0\right)\to\rho(p,p,0)$ as
$q\to0$. Inserting Eq.~\eqref{eq:generic_initial_kummer_scaling} into
Eq.~\eqref{eq:generic_initial_characteristic_exact}, we then obtain
\begin{equation}
\label{eq:generic_initial_characteristic_asymptotic}
\widetilde P_r(q,t)
\approx
\int_{-\infty}^{\infty}
\mathrm{d}p\,
\rho(p,p,0)
\mathrm{e}^{-\mathrm{i}q t_{\mathrm{eff}}(t)v(p)}.
\end{equation}
We now invert the Fourier transform in
Eq.~\eqref{eq:generic_initial_characteristic_asymptotic}. This gives
\begin{align}
P_r(x,t)
&\approx
\frac{1}{2\pi}
\int_{-\infty}^{\infty}
\mathrm{d}q\,
\mathrm{e}^{\mathrm{i}qx}
\int_{-\infty}^{\infty}
\mathrm{d}p\,
\rho(p,p,0)
\mathrm{e}^{-\mathrm{i}q t_{\mathrm{eff}}(t)v(p)}
\nonumber\\
&=
\int_{-\infty}^{\infty}
\mathrm{d}p\,
\rho(p,p,0)
\frac{1}{2\pi}
\int_{-\infty}^{\infty}
\mathrm{d}q\,
\mathrm{e}^{\mathrm{i}q[x-t_{\mathrm{eff}}(t)v(p)]}
\nonumber\\
&=
\int_{-\infty}^{\infty}
\mathrm{d}p\,
\rho(p,p,0)
\delta\left(
x-t_{\mathrm{eff}}(t)v(p)
\right).
\label{eq:generic_position_delta_form}
\end{align}
Using the identity
$\delta\left(x-t_{\mathrm{eff}}v(p)\right)
=\frac{1}{t_{\mathrm{eff}}}
\delta\left(\frac{x}{t_{\mathrm{eff}}}-v(p)\right)$,
Eq.~\eqref{eq:generic_position_delta_form} becomes
\begin{equation}
P_r(x,t)
\approx
\frac{1}{t_{\mathrm{eff}}(t)}
\int_{-\infty}^{\infty}
\mathrm{d}p\,
\rho(p,p,0)
\delta\left(
\frac{x}{t_{\mathrm{eff}}(t)}-v(p)
\right).
\end{equation}
We now define the group-velocity distribution induced by the diagonal
momentum distribution of the initial state $\rho(p,p,0)$ as
\begin{equation}
\label{eq:generic_initial_velocity_distribution}
\mathcal V_0(v)
=
\int_{-\infty}^{\infty}
\mathrm{d}p\,
\rho(p,p,0)
\delta\left(
v-\varepsilon'(p)
\right)\,,
\end{equation}
where $\varepsilon'(p)= \frac{d\varepsilon(p)}{dp}$. Using Eq.~\eqref{eq:generic_initial_velocity_distribution}, we finally
obtain
\begin{equation}
\label{eq:generic_initial_general_scaling_result}
P_r(x,t)
\approx
\frac{1}{t_{\mathrm{eff}}(t)}
\mathcal V_0\left(
\frac{x}{t_{\mathrm{eff}}(t)}
\right).
\end{equation}
Since $t_{\mathrm{eff}}(t)=\log(rt)/r$, this result can equivalently be
written as
\begin{equation}
P_r(x,t)
\approx
\frac{r}{\log(rt)}
\mathcal V_0\left(
\frac{rx}{\log(rt)}
\right).
\end{equation}
Thus, the characteristic spatial scale is completely universal and it grows logarithmically as
$\log(rt)/r$, while the shape of the asymptotic distribution is non-universal and given
by the initial group-velocity distribution in Eq. \eqref{eq:generic_initial_velocity_distribution}.

\par\vspace{1.5\baselineskip}

\end{document}